\documentclass[hidelinks,onefignum,onetabnum]{siamart220329}

\usepackage{lipsum}
\usepackage{amsfonts}
\usepackage{graphicx}
\usepackage{epstopdf}
\usepackage{algorithmic}
\usepackage{orcidlink}
\usepackage{amsmath,amssymb,amsfonts}
\usepackage{mathtools}
\usepackage{bm}
\usepackage{caption}
\usepackage{subcaption}
\usepackage{tabularx}
\usepackage{siunitx}
\ifpdf
  \DeclareGraphicsExtensions{.eps,.pdf,.png,.jpg}
\else
  \DeclareGraphicsExtensions{.eps}
\fi

\newsiamremark{remark}{Remark}
\newsiamremark{hypothesis}{Hypothesis}
\crefname{hypothesis}{Hypothesis}{Hypotheses}
\newsiamthm{claim}{Claim}

\headers{Block preconditioners for flow rate conditions}{L. Crugnola, and C. Vergara}

\title{Inexact block LU preconditioners for incompressible fluids with flow rate conditions}

\author{ \orcidlink{0000-0001-5702-8440}Luca Crugnola\thanks{LaBS, Department of Chemistry, Materials and Chemical Engineering “Giulio Natta”, Politecnico di Milano, Piazza Leonardo da Vinci 32, Milan, 20133, Italy (\email{luca.crugnola@polimi.it}).}
\and \orcidlink{0000-0001-9872-5410}Christian Vergara\thanks{LaBS, Department of Chemistry, Materials and Chemical Engineering “Giulio Natta”, Politecnico di Milano, Piazza Leonardo da Vinci 32, Milan, 20133, Italy (\email{christian.vergara@polimi.it}).}}

\usepackage{amsopn}

\ifpdf
\hypersetup{
  pdftitle={Inexact block LU preconditioners with flow rate conditions},
  pdfauthor={L. Crugnola, and C. Vergara}
}
\fi

\begin{document}

\maketitle

\begin{abstract}
When studying the dynamics of incompressible fluids in bounded domains the only available data often provide average flow rate conditions on portions of the domain's boundary. In engineering applications a common practice to complete these conditions is to prescribe a Dirichlet condition by assuming a-priori a spatial profile for the velocity field. However, this strongly influence the accuracy of the numerical solution. A more mathematically sound approach is to prescribe the flow rate conditions using Lagrange multipliers, resulting in an augmented weak formulation of the Navier-Stokes problem. 

In this paper we start from the SIMPLE preconditioner, introduced so far for the standard Navier-Stokes equations, and we derive two preconditioners for the monolithic solution of the augmented problem. This can be useful in complex applications where splitting the computation of the velocity/pressure and Lagrange multipliers numerical solutions can be very expensive. In particular, we investigate the numerical performance of the preconditioners in both idealized and real-life scenarios. Finally, we highlight the advantages of treating flow rate conditions with a Lagrange multipliers approach instead of prescribing a Dirichlet condition.
\end{abstract}

\begin{keywords}
Flow rate conditions, Lagrange multipliers, SIMPLE preconditioner, Incompressible Navier-Stokes equations, Computational Fluid-Dynamics
\end{keywords}

\section{Introduction}

Incompressible Navier-Stokes equations are commonly used in engineering applications to study the dynamics of viscous fluids. Given a bounded domain $\Omega\subset\mathbb{R}^3$, classical boundary conditions impose three (one for each spatial dimension) point-wise data on the domain's boundary $\partial\Omega$, typically prescribing the components of the velocity field (Dirichlet condition), of the Cauchy normal stress (Neumann condition), or a combination of the two (Robin condition). 

On the other hand, in some applications the only information available provides average conditions which are not enough to close the problem and thus need to be completed \cite{conca1995navier}. We will refer to them as \textit{defective} boundary conditions. Some examples arise in blood-dynamics simulations, where clinical measures often provide information on either the \textit{flow rate} or the \textit{mean pressure} \cite{quarteroni2016geometric,veneziani2005flow}, and flow in pipes simulations, where sensors measure the fluid flow rate \cite{bonakdari2011influence,pittard2004experimental}. Moreover, in the \textit{geometric multiscale approach} \cite{quarteroni2001coupling}, where mathematical models with different spatial dimensions are coupled, the information provided by the lower-dimensional problem is not enough to close the higher-dimensional one \cite{blanco2009potentialities,leiva2011partitioned}.

Regarding defective flow rate conditions, a widely employed strategy to close the problem, thanks to its practicality, is to prescribe a Dirichlet condition by assuming a-priori a spatial profile for the velocity field. In this case, the computational domain is often extended to reduce the impact of the chosen profile on the accuracy of the numerical solution. A more mathematically sound approach was proposed in \cite{heywood1996artificial}, where the authors introduced a suitable variational formulation of the problem which includes the given flow rate data. However, this approach requires the definition of non-standard finite-dimensional subspaces which makes it problematic to implement in practice \cite{veneziani2007approximate}. Hence, several alternative approaches have been proposed in literature \cite{formaggia2012prescription} based on Lagrange multipliers \cite{formaggia2003numerical,veneziani2005flow}, control theory \cite{formaggia2008new,lee2011optimal}, penalization technique \cite{zunino2009numerical} and the Nitsche method \cite{juntunen2009nitsche,vergara2011nitsche}. 

The Lagrange multipliers approach was proposed in \cite{formaggia2003numerical} and applied for a quasi-Newtonian Stokes problem \cite{ervin2007numerical}, in a fluid-structure interaction framework \cite{formaggia2010flow} and in practical hemodynamic problems with patient-specific geometries \cite{vergara2012influence,crugnola2024computational}. In this approach, flow rate boundary conditions are considered as a constraint for the solution and enforced using Lagrange multipliers, resulting in an augmented weak formulation of the Navier-Stokes problem. The problem is closed by assuming that, on the considered portion of the domain's boundary, the Cauchy normal stress has zero tangential components and its normal component is constant in space. The resulting augmented problem can be solved by splitting the computation of the velocity/pressure fields and of the Lagrange multipliers in order to resort to available standard solvers for the solution of the Navier-Stokes step \cite{veneziani2005flow,veneziani2007approximate}. However, iterative procedures based on this splitting can be very expensive in complex applications. In the recent work \cite{hirschvogel2024effective}, the authors proposed monolithic block preconditioners based on inexact factorizations to deal with defective conditions arising from the coupling with lumped parameter models.

In order to efficiently solve the augmented Finite Elements system, in this paper we consider a monolithic strategy and we propose a suitable block preconditioner for its efficient solution. Specifically, we start from the SIMPLE iterative solution strategy \cite{patankar1972calculation}, studied as a preconditioner in \cite{pernice2001multigrid,li2004eigenvalue,elman2006block,elman2008taxonomy}, and we extend it to the augmented flow rate defective case. To test the effectiveness of our proposal, we present several numerical results both in idealized and real-life scenarios.

The outline of the paper is as follows: we provide a review of the augmented Navier-Stokes flow rate problem (Section \ref{sec:aug-NS-prob}) and of the SIMPLE preconditioner (Section \ref{sec:SIMPLE-prec-rev}); we propose an extension of the SIMPLE preconditioner (Section \ref{sec:ext-SIMPLE-prec}), discussing its formulation (Section \ref{subsec:aug-SIMPLE}) and introducing an efficient variant used in practice (Section \ref{subsec:aug-aSIMPLE}); finally we provide some numerical results (Section \ref{sec:num-results}), presenting the setting of our numerical experiments which involves a trivial extension of the SIMPLE preconditioner (Section \ref{subsec:num-exp-set}), testing the performance of the proposed preconditioners for a varying number of Lagrange multipliers (Section \ref{subsec:templ-geom}) and in real-life hemodynamic scenarios (Section \ref{subsec:real-life-appl}), and highlighting the advantages of treating defective flow rate condition with a Lagrange multipliers approach (Section \ref{subsec:Def-vs-Dir}).

\section{Review of the augmented flow rate problem}
\label{sec:aug-NS-prob}

\begin{figure}[hbt]
    \centering
    \includegraphics[width=0.5\textwidth]{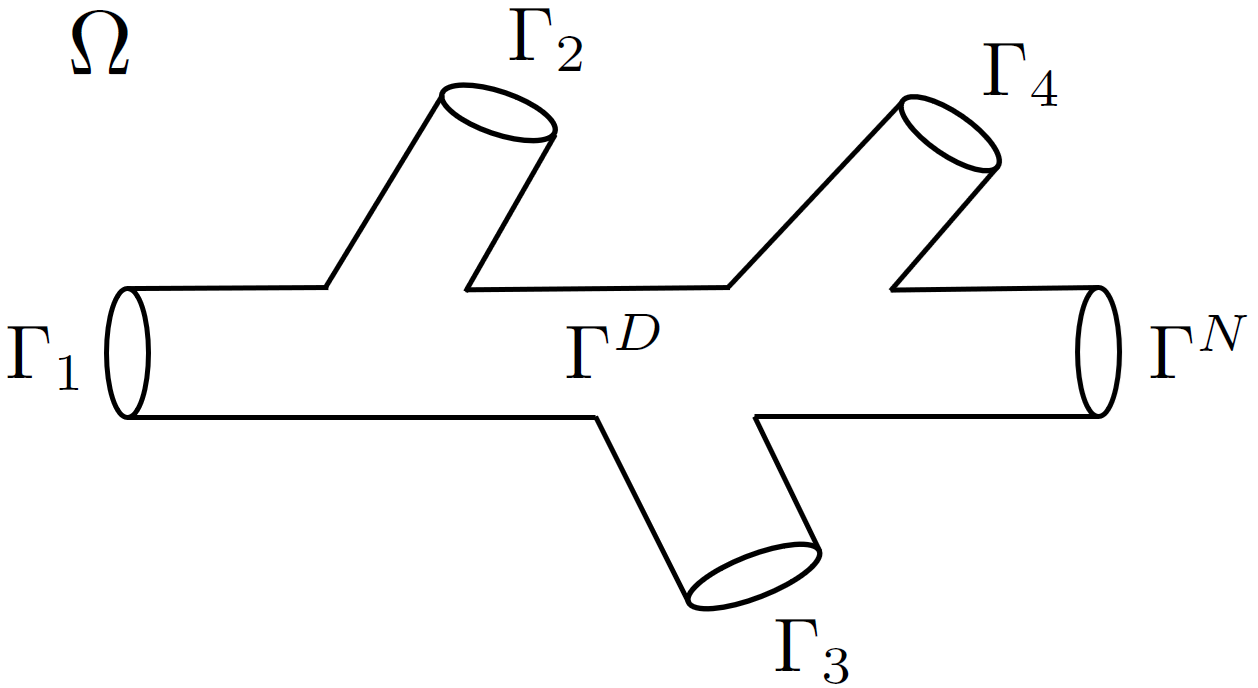}
    \caption{Sketch of a domain $\Omega$, with the partition of its boundary. In this example $m=4$.}\label{fig:domWithBound_sketch}
\end{figure}

We are interested in solving the incompressible unsteady Navier-Stokes equations for a homogeneous Newtonian fluid in a bounded domain $\Omega\subset\mathbb{R}^3$. In particular, we consider the following boundary conditions:
\begin{subequations}\label{eq:BCs}
	\begin{align}
		&\textbf{u} = \textbf{0}\ &&\text{on }\Gamma^{D}\label{eq:Dir-cond} \\
		&-p\textbf{n}+\nu\frac{\partial\textbf{u}}{\partial\textbf{n}} = \bm{0}\ &&\text{on }\Gamma^{N}\label{eq:Neum-cond} \\
		&\int_{\Gamma_i}{\textbf{u}\cdot\textbf{n}\ d\gamma}=Q_i\ &&\text{on }\Gamma_{i},\ i=1,\ldots,m\label{eq:Def-cond}
	\end{align}
\end{subequations}
where $\textbf{u}$ and $p$ are the fluid velocity and pressure, respectively, $\nu$ is the kinematic viscosity, $\textbf{n}$ is the outward unit normal vector on $\partial\Omega$ and $\partial\Omega=\Gamma^D\cup\Gamma^N$$\cup\left(\bigcup_{i=1}^m{\Gamma_i}\right)$ (see Figure \ref{fig:domWithBound_sketch}). Notice that, for the sake of simplicity, we assume that the fluid density is equal to $1\,g/cm^3$. Moreover, we are considering homogeneous Dirichlet and Neumann conditions; all the findings of this paper could be easily extended to the non-homogeneous case as well. Equation \eqref{eq:Def-cond} expresses defective boundary conditions, imposing time-dependent flow rates $Q_i=Q_i(t)$ on sections $\Gamma_i$, $i=1,\ldots,m$. We assume $\Gamma^N\neq\emptyset$ (which is the case of interest in our numerical experiments), so that the incompressibility constraint is satisfied without any restriction on the choice of the flow rates $Q_i$ \cite{formaggia2003numerical}.

Let us introduce the following functional space:
\begin{equation*}
	V=\left\{\textbf{v}\in [H^1(\Omega)]^3
 : \textbf{v}=\textbf{0}\ \text{on } \Gamma^D\right\}.
\end{equation*}
Then, the augmented variational formulation, corresponding to boundary conditions \eqref{eq:BCs}, as proposed in \cite{formaggia2003numerical}, reads:\\
\noindent Given $\textbf{f}\in L^2(\Omega)$ for almost all $t\in(0,T]$, $Q_i:[0,T]\to\mathbb{R}$, $i=1,\ldots,m$, and $\textbf{u}_0\in V$, find $\textbf{u}\in V$ and $p\in L^2(\Omega)$, for almost all $t\in(0,T]$, and functions of time $\lambda_i:[0,T]\to\mathbb{R}$, $i=1,\ldots,m$, such that, for all $\textbf{v}\in V$ and $q\in L^2(\Omega)$,
\begin{align}\label{eq:aug-var-form}
        \nonumber\displaystyle &\left(\frac{\partial\textbf{u}}{\partial t}+\textbf{u}\cdot\nabla\textbf{u},\textbf{v}\right) + \nu\left(\nabla\textbf{u},\nabla\textbf{v}\right) - \left(p,\nabla\cdot\textbf{v}\right) + \sum\limits_{i=1}^{m}{\lambda_i\int_{\Gamma_i}{\textbf{v}\cdot\textbf{n}\ d\gamma}} = \left(\textbf{f},\textbf{v}\right),\\
        \displaystyle &\left(q,\nabla\cdot\textbf{u}\right)=0,\\
        \nonumber\displaystyle &\int_{\Gamma_i}{\textbf{u}\cdot\textbf{n}\ d\gamma}=Q_i,\hspace{0.3cm}i=1,...,m,
\end{align}
with $\textbf{u}=\textbf{u}_0$ for $t=0$. 

In formulation \eqref{eq:aug-var-form}, the expression $\left(\cdot\, ,\cdot\right)$ stands for the $L^2$ scalar product in $\Omega$, for scalars, vectors and tensors, whereas $\lambda_i$ are the Lagrange multipliers introduced to enforce the defective flow rate boundary conditions \eqref{eq:Def-cond}. 

We consider a Galerkin approximation of formulation \eqref{eq:aug-var-form} by introducing the finite-dimensional spaces $V_h\subset V$, with basis functions $\left\{\bm{\psi}_j\right\}_{j=1}^{N_u}$, and $M_h\subset L^2(\Omega)$, with basis functions $\left\{\xi_\ell\right\}_{\ell=1}^{N_p}$. Moreover, we consider a first order backward difference formula discretization of the time derivative with a semi-implicit treatment of the convective term. Then, the augmented algebraic monolithic system associated to formulation \eqref{eq:aug-var-form}, accounting for possible stabilization terms, reads:
\begin{equation}\label{eq:aug-alg-sys}
	\begin{bmatrix}
	K & B^T & \Phi^T \\
	-B & S & 0 \\
	\Phi & 0 & 0
	\end{bmatrix}
	\begin{bmatrix}
	\bm{U} \\
	\bm{P} \\
	\bm{\Lambda}
	\end{bmatrix}
	=
	\begin{bmatrix}
	{\bm{F}} \\
	\bm{0} \\
	\bm{Q}
	\end{bmatrix}
\end{equation}
where $\bm{U}\in\mathbb{R}^{N_u}$, $\bm{P}\in\mathbb{R}^{N_p}$ and $\bm{\Lambda}\in\mathbb{R}^{m}$ are the solution vectors for the velocity, pressure and Lagrange multipliers, respectively, and the current time-step $n+1$ is understood. Moreover, $K=\frac{1}{\Delta t}M+A+C(\bm{U}^n)$, where $\Delta t$ is the time-step, $M$ is the velocity mass matrix, $A$ is the stiffness matrix and $C(\bm{U}^n)$ is the matrix associated to the convective term, with $\textbf{U}^n\simeq\textbf{U}(t^n), \, t^n=n\Delta t, \, n=0,1,\ldots,N_T$. Finally, $[B]_{\ell j}=-(\xi_\ell,\nabla\cdot\bm{\psi}_j)$, $\ell=1,\ldots,N_p$, $j=1,\ldots,N_u$, $[\Phi]_{ij}=\int_{\Gamma_i}{\bm{\psi}_j\cdot\textbf{n}\ d\gamma}$, $i=1,\ldots,m$ $j=1,\ldots,N_u$, $S$ accounts for stabilization terms (such as SUPG-PSPG \cite{tezduyar2002calculation}), $\left[{\bm{F}}\right]_j = (\textbf{f},\bm{\psi}_j) + \frac{1}{\Delta t}\left[M\bm{U}^n\right]_j$, $j=1,\ldots,N_u$, and $[\bm{Q}]_i=Q_i$, $i=1,\ldots,m$.

\section{Review of the SIMPLE preconditioner}
\label{sec:SIMPLE-prec-rev}

Let us consider the algebraic system arising from the discretization of Navier-Stokes equations with classical boundary conditions, accounting for possible stabilization terms:
\begin{equation}\label{eq:NS-alg-sys-2x2}
	A\bm{x}
	\coloneq
	\begin{bmatrix}
		K & B^T \\
		-B & S
	\end{bmatrix}
	\begin{bmatrix}
		\bm{U} \\
		\bm{P}
	\end{bmatrix}
	= 
	\begin{bmatrix}
		\bm{F} \\
		\bm{0}
	\end{bmatrix}
        \eqcolon
	\bm{b}
\end{equation}
where each matrix and vector in system \eqref{eq:NS-alg-sys-2x2} was defined at the end of Section \ref{sec:aug-NS-prob}.

The SIMPLE algorithm was proposed in \cite{patankar1972calculation} as an iterative method to solve system \eqref{eq:NS-alg-sys-2x2}. We report here a SIMPLE-like solver algorithm which first solves for a compressible intermediate velocity, then solves for the pressure and finally compute a velocity correction in order to impose mass-continuity: 
\vspace{0.2cm}\\
\noindent \textit{SIMPLE-like solver algorithm}:\vspace{0.1cm}
\begin{enumerate}
	\item Solve for the intermediate velocity: $K\widetilde{\bm{U}}=\bm{F}$,\vspace{0.1cm}
	\item Solve for the pressure: $\Sigma\bm{P}=B\widetilde{\bm{U}}$,\vspace{0.1cm}
	\item Compute the velocity correction: $\bm{U} = \widetilde{\bm{U}} - D^{-1}B^T\bm{ P}$,\vspace{0.1cm}
\end{enumerate}
where $\Sigma=BD^{-1}B^T+S$ is an approximation of the pressure Schur complement $BK^{-1}B^T+S$, with $D=diag(K)$. Note that using a diagonal approximation of $K$ makes it possible to explicitly invert the matrix and thus to assemble the approximate Schur complement.

This algorithm can be seen as an approximate solution of problem \eqref{eq:NS-alg-sys-2x2}, obtained by using the following inexact block LU factorization of $A$:
\begin{equation}\label{eq:SIMPLEprec}
	P_S
	=
	\begin{bmatrix}
		K & 0 \\
		-B & \Sigma
	\end{bmatrix}
	\begin{bmatrix}
		I & D^{-1}B^T \\
		0 & I
	\end{bmatrix}.
\end{equation}
On the other hand, one can use $P_S$ as a preconditioner to solve problem \eqref{eq:NS-alg-sys-2x2} with a Krylov subspace method, such as GMRES \cite{saad2003iterative}. So that, the computation of the preconditioned residual at the  $k$-th iteration of the Krylov method reads
\begin{equation*}
	P_S\bm{z}^{(k)}
	=
	\bm{b} - A
	\bm{x}^{(k)}
	\eqcolon
	\bm{r}^{(k)},
\end{equation*}
that is:
\begin{align*}
	&\begin{cases}
		K\bm{y}_1^{(k)} = \bm{r}_1^{(k)}, \\
		\Sigma \bm{y}_2^{(k)} = B \bm{y}_1^{(k)} + \bm{r}_2^{(k)},
	\end{cases}\\
	&\begin{cases}
		\bm{z}_1^{(k)} = \bm{y}_1^{(k)} - D^{-1}B^T \bm{z}_2^{(k)}, \\
		\bm{z}_2^{(k)} = \bm{y}_2^{(k)},
	\end{cases}
\end{align*}
where given $\textbf{w}\in\mathbb{R}^{N_U+N_P}$ we indicate by $\textbf{w}_1\in\mathbb{R}^{N_U}$ the first $N_U$ components and by $\textbf{w}_2\in\mathbb{R}^{N_P}$ the last $N_P$ components and $\bm{y}^{(k)}$ is an intermediate vector.

\begin{remark}\label{rem:general-LU-2x2}
    Several approximate solution methods and corresponding preconditioners for system \eqref{eq:NS-alg-sys-2x2} are based on an inexact block LU factorization of the system matrix. We report here a general form of the factorization, as proposed in \cite{quarteroni2000factorization}:
    \begin{equation*}
        \begin{bmatrix}
		K & 0 \\
		-B & BH_1B^T+S
	\end{bmatrix}
	\begin{bmatrix}
		I & H_2B^T \\
		0 & I
	\end{bmatrix},
    \end{equation*}
\end{remark}
where $H_1$ and $H_2$ are approximations of $K^{-1}$. Some notable examples are SIMPLE ($H_1=H_2=diag(K)^{-1}$), Chorin-Temam ($H_1=H_2=\Delta t M^{-1}$) and Yosida ($H_1=\Delta t M^{-1}$, $H_2=K^{-1}$), where $M$ is the velocity mass matrix.

\section{Extension of the SIMPLE preconditioner to the flow rate augmented problem}
\label{sec:ext-SIMPLE-prec} 

In this section we present the main novelty of this work which is the extension, based on an inexact factorization, of the standard SIMPLE preconditioner \eqref{eq:SIMPLEprec} to the augmented problem \eqref{eq:aug-alg-sys}. We stress that this extension is in principle not immediate due to the $3\times3$ nature of the matrix in \eqref{eq:aug-alg-sys} (vs the $2\times2$ nature of the standard NS matrix), which requires specific and suitable choices of the inexact factors. 

\subsection{Formulation of the preconditioner}\label{subsec:aug-SIMPLE}

According to the augmented algebraic system \eqref{eq:aug-alg-sys}, we set:
\begin{equation}\label{eq:aug-alg-matrix}
	A^{aug}
	=
	\begin{bmatrix}
		K & B^T & \Phi^T \\
		-B & S & 0 \\
		\Phi & 0 & 0
	\end{bmatrix}.
\end{equation}
In analogy with \eqref{eq:SIMPLEprec} and according to the notation of Section \ref{sec:SIMPLE-prec-rev}, we propose the following inexact block LU factorization of matrix \eqref{eq:aug-alg-matrix}:
\begin{equation}\label{eq:aug-SIMPLE-prec}
	P_{S}^{aug} 
	=
	\begin{bmatrix}
		K & 0 & 0 \\
		-B & \Sigma & 0 \\
		\Phi & -\Sigma_{\Lambda P} & L_{33}
	\end{bmatrix}
	\begin{bmatrix}
		I & D^{-1}B^T & D^{-1}\Phi^T \\
		0 & I & W^{-1}\Sigma_{P\Lambda} \\
		0 & 0 & I
	\end{bmatrix},
\end{equation}
where $\Sigma_{\Lambda P}=\Phi D^{-1}B^T$, $L_{33}=\Sigma_{\Lambda P}W^{-1}\Sigma_{P\Lambda}-\Sigma_{\Lambda}$, $\Sigma_{P\Lambda}=BD^{-1}\Phi^T$, $\Sigma_{\Lambda}=\Phi D^{-1}\Phi^T$ and $W = diag(\Sigma)$.
In particular, as happens for the standard case, all the matrices appearing in this preconditioner can be explicitly assembled, thanks to the use of the diagonal approximations $D$ and $W$.

In what follows we detail the algebraic steps needed at the $k$-th iteration of a Krylov method to compute the preconditioned residual associated to the augmented SIMPLE matrix \eqref{eq:aug-SIMPLE-prec}:
\begin{equation*}\label{eqs:KrylovIterAug}
	P_{S}^{aug}\bm{z}^{(k)}
	=
	\bm{r}^{(k)},
\end{equation*}
that is:
\begin{align}\label{eq:aug-SIMPLE}
	&\begin{cases}
		K\bm{y}_1^{(k)} = \bm{r}_1^{(k)}, \\
		\Sigma\bm{y}_2^{(k)} = \bm{r}_2^{(k)} +B\bm{y}_1^{(k)}, \\
		{L}_{33}\bm{y}_3^{(k)} = \bm{r}_3^{(k)} -\Phi\bm{y}_1^{(k)} +\Sigma_{\Lambda P}\bm{y}_2^{(k)},
	\end{cases} \\
	\nonumber&\begin{cases}
		\bm{z}_1^{(k)} = \bm{y}_1^{(k)} -D^{-1}B^T\bm{z}_2^{(k)} -D^{-1}\Phi^T\bm{z}_3^{(k)},  \\
		\bm{z}_2^{(k)} = \bm{y}_2^{(k)} -W^{-1}\Sigma_{P\Lambda}\bm{z}_3^{(k)}, \\
		\bm{z}_3^{(k)} = \bm{y}_3^{(k)},
	\end{cases}
\end{align}
where, given $\textbf{w}\in\mathbb{R}^{N_U+N_P+m}$, we indicate by $\textbf{w}_1\in\mathbb{R}^{N_U}$ the first $N_U$ components associated to the velocity d.o.f.'s, by $\textbf{w}_2\in\mathbb{R}^{N_P}$ the following $N_P$ components associated to the pressure d.o.f.'s, and by $\textbf{w}_3\in\mathbb{R}^{m}$ the last $m$ components associated to the Lagrange multipliers d.o.f.'s. 

\begin{remark}
    We notice that in \cite{hirschvogel2024effective} the authors just very recently proposed a monolithic preconditioner for a similar $3\times3$ block system arising from the discretization of the 3D-0D problem, associated to the coupling with a lumped parameter model. However, in their case, the last block row accounts for the coupling constraint equations, resulting in a form of the $3\times3$ system's block matrix and of the corresponding inexact factorization which are different from ours. 
\end{remark}

\begin{remark}
    Notice that our preconditioner originates from the exact block LU factorization associated to the augmented matrix \eqref{eq:aug-alg-matrix}:
    \begin{align}\label{eq:aug-exact-LU}
	&\nonumber L=
        \begin{bmatrix}
		K & 0 & 0 \\
		-B & BK^{-1}B^T+S & 0 \\
		\Phi & -\Phi K^{-1}B^T & \Phi K^{-1}B^T(BK^{-1}B^T+S)^{-1}BK^{-1}\Phi^T-\Phi K^{-1}\Phi^T
	\end{bmatrix}, \\
        &U = 
        \begin{bmatrix}
		I & K^{-1}B^T & K^{-1}\Phi^T \\
		0 & I & (BK^{-1}B^T+S)^{-1}BK^{-1}\Phi^T \\
		0 & 0 & I
	\end{bmatrix}.
\end{align}
    In particular, the augmented SIMPLE preconditioner \eqref{eq:aug-SIMPLE-prec} can be obtained by using $D^{-1}$ and $W^{-1}$ as an approximation of $K^{-1}$ and $\Sigma^{-1}$, leading to a coherent extension of the standard SIMPLE preconditioner \eqref{eq:SIMPLEprec}. According to Remark \ref{rem:general-LU-2x2}, other preconditioners based on an inexact block LU factorization of the Navier Stokes matrix (e.g. Chorin-Temam, Yosida) can be extended to the augmented case by considering different approximations of the inverse matrices. 
\end{remark}

\begin{remark}
    Also for the case of the extension to the augmented flow rate problem, it is possible to write the SIMPLE-like solver algorithm associated to the preconditioner \eqref{eq:aug-SIMPLE-prec}.\vspace{0.2cm}\\
\noindent \textit{Augmented SIMPLE-like solver algorithm}:\vspace{0.1cm}
\begin{enumerate}
	\item Solve for the intermediate velocity: $K\widetilde{\bm{U}}=\bm{F}$,\vspace{0.1cm}
	\item Solve for the intermediate pressure: $\Sigma\widetilde{\bm{P}}=B\widetilde{\bm{U}},$\vspace{0.1cm}
	\item Solve for the Lagrange multipliers: $L_{33}\bm{\Lambda}=\bm{Q}+\Sigma_{\Lambda P}\widetilde{\bm{P}}-\Phi\widetilde{\bm{U}}$,\vspace{0.1cm}
	\item Compute the pressure correction: $\bm{P} = \widetilde{\bm{P}} - W^{-1}\Sigma_{P\Lambda}\bm{\Lambda}$,\vspace{0.1cm}
	\item Compute the velocity correction: $\bm{U} = \widetilde{\bm{U}} - D^{-1}B^T\bm{ P} - D^{-1}\Phi^T\bm{\Lambda}$,\vspace{0.2cm}
\end{enumerate}
Similarly to what is done in \cite{elman2008taxonomy}, from definition \eqref{eq:aug-SIMPLE-prec} it is easy to show that
\begin{equation*}
	P_S^{aug}
	=
	\begin{bmatrix}
		K & KD^{-1}B^T & KD^{-1}\Phi^T \\
		-B & S & -(I-\Sigma W^{-1}) \Sigma_{P\Lambda} \\
		\Phi & 0 & 0
	\end{bmatrix},
\end{equation*}
and that we have the following error:
\begin{equation*}
	A^{aug}-P_S^{aug}
	=
	\begin{bmatrix}
		0 & (I-KD^{-1})B^T & (I-KD^{-1})\Phi^T \\
		0 & 0 & (I-\Sigma W^{-1}) \Sigma_{P\Lambda} \\
		0 & 0 & 0 \\
	\end{bmatrix}.
\end{equation*}
Hence, this augmented SIMPLE-like solver introduces an error in both the momentum and continuity equations, thus it loses the "mass-preserving" property which holds true for the standard case. On the other hand, the defective flow rate boundary conditions are imposed exactly. Finally, we observe that if matrices $K$ and $\Sigma$ are well approximated by their diagonal, then $P_S^{aug}$ is a good approximation of $A^{aug}$.
\end{remark}

\subsection{An efficient variant of the augmented SIMPLE preconditioner}\label{subsec:aug-aSIMPLE}

We observe that the solution of the L-step \eqref{eq:aug-SIMPLE} requires to solve linear systems in $K$, $\Sigma$ and $L_{33}$, which may present in general some difficulties due to the ill-conditioning of matrix $K$ and the dense nature of matrices $\Sigma$ and $L_{33}$. Hence, inspired by the aSIMPLE preconditioner presented in \cite{deparis2014parallel}, in our numerical experiments we substitute matrices $K$, $\Sigma$ and $L_{33}$ in $P_S^{aug}$ \eqref{eq:aug-SIMPLE-prec} with approximations $\widehat{K}$, $\widehat{\Sigma}$ and $\widehat{L}_{33}$ based on the Algebraic Multigrid method. Specifically, we use the Chebyshev pre-smoothing and the KLU solver for coarse level \cite{zhang2010explicit}. 

\section{Numerical results}
\label{sec:num-results}

\subsection{Setting of the numerical experiments}\label{subsec:num-exp-set}
For the numerical solution of the augmented monolithic system \eqref{eq:aug-alg-sys}, we aim to investigate the performance of the augmented aSIMPLE preconditioner (introduced in Section \ref{subsec:aug-aSIMPLE}):
\begin{equation}\label{eq:aug-aSIMPLE-inexact}
        P_{aS}^{aug}
        =
	\begin{bmatrix}
		\widehat{K} & 0 & 0 \\
		-B & \widehat{\Sigma} & 0 \\
		\Phi & -\Sigma_{\Lambda P} & \widehat{L}_{33}
	\end{bmatrix}
	\begin{bmatrix}
		I & D^{-1}B^T & D^{-1}\Phi^T \\
		0 & I & W^{-1}\Sigma_{P\Lambda} \\
		0 & 0 & I
	\end{bmatrix},
\end{equation}
by comparing it with a trivially implementable extension of the standard aSIMPLE preconditioner \cite{deparis2014parallel}, built as follows. We observe that in system \eqref{eq:aug-alg-sys} the dimension of the Lagrange multipliers block ($m$) is given by the number of defective flow rate conditions prescribed with Lagrange multipliers, which is usually significantly smaller than the dimension of the velocity ($N^U$) and pressure ($N^P$) blocks. Thus, one could be tempted to extend to the augmented case the standard aSIMPLE preconditioner (obtained from \eqref{eq:SIMPLEprec} with suitable approximations of $K$ and $\Sigma$ \cite{deparis2014parallel}) by simply adding identity blocks on the diagonals:
\begin{equation}\label{eq:aug-aSIMPLE-identity}
        P_{aS-I}^{aug}
        =
	\begin{bmatrix}
		\widehat{K} & 0 & 0 \\
		-B & \widehat{\Sigma} & 0 \\
		0 & 0 & I
	\end{bmatrix}
	\begin{bmatrix}
		I & D^{-1}B^T & 0 \\
		0 & I & 0 \\
		0 & 0 & I
	\end{bmatrix}.
\end{equation}
In the following we will refer to the complete preconditioner $P_{aS}^{aug}$ \eqref{eq:aug-aSIMPLE-inexact} as \textit{aug-aS} and to the simplified preconditioner $P_{aS-I}^{aug}$ \eqref{eq:aug-aSIMPLE-identity} as \textit{aug-aS-I}.

Notice that in the numerical experiments we employ piecewise linear Finite Elements for the velocity and pressure fields with a SUPG/PSPG \cite{tezduyar2002calculation} stabilization scheme, considering a kinematic viscosity $\nu=3.3\times10^{-6}\ m^2/s$. Moreover, we solve the monolithic augmented system \eqref{eq:aug-alg-sys} using a preconditioned GMRES iterative solver with both preconditioners aug-aS and aug-aS-I. Finally, all the numerical experiments are performed using the multi-physics high-performance library $\mathrm{life}^\mathrm{x}$ \cite{africa2022lifexcore,africa2024lifexcfd} (\url{https://lifex.gitlab.io}, \url{https://doi.org/10.5281/zenodo.7852088}), developed at MOX, Dipartimento di Matematica, with the collaboration of LaBS, Dipartimento di Chimica, Materiali ed Ingengneria Chimica (both at Politecnico di Milano). 

We consider the following tests:
\begin{itemize}
    \item[-] Test I: idealized geometry to assess the influence of the number of Lagrange multipliers;
    \item[-] Test II: simulation in a stenotic carotid artery ($m=2$);
    \item[-] Test III: simulation in coronary arteries in presence of a double bypass ($m=3$);
    \item[-] Test IV: simulation in an aorta with bicuspid aortic valve ($m=4$). 
\end{itemize} 

\subsection{Test I - Varying the number of Lagrange multipliers}
\label{subsec:templ-geom}

\begin{figure}[hbt]
    \centering
    \begin{subfigure}[b]{0.32\textwidth}
        \centering
        \includegraphics[width=\linewidth]{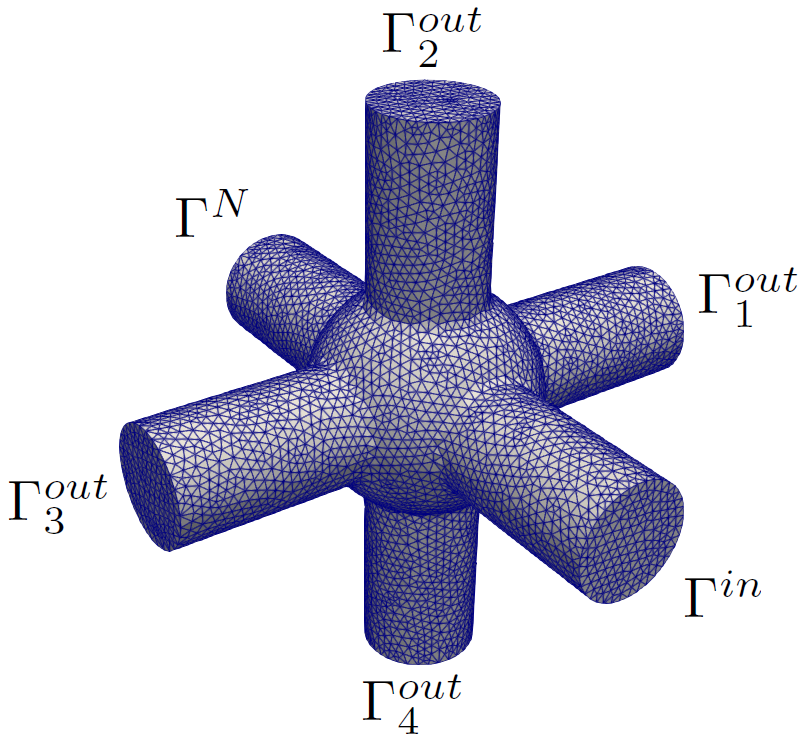}
        \caption{}\label{fig::starShape_domain}
    \end{subfigure}
    \hspace{0.5cm}
    \begin{subfigure}[b]{0.4\textwidth}
        \centering
        \includegraphics[width=\linewidth]{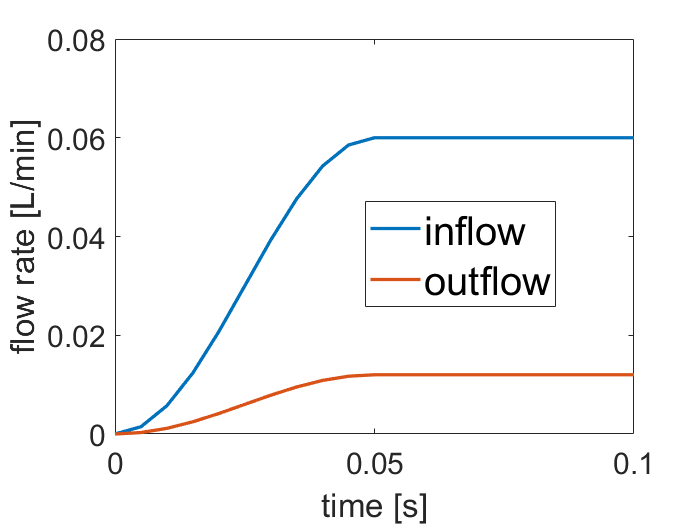}
        \caption{}\label{fig::ramp_in-out}
    \end{subfigure}
    \caption{(a) Star-shape domain comprising six cylinders and a sphere ($h\approx0.2\ mm$). (b) Time-evolving flow rates prescribed at the inlet section $\Gamma^{in}$ (inflow) and at each outlet section $\Gamma^{out}_i$, $i=1,\ldots,4$ (outflow). Test I.}\label{fig::starShape_inflow-outflow}
\end{figure}

We are interested in assessing the numerical performance of preconditioners aug-aS and aug-aS-I for an increasing number of flow rate conditions prescribed with Lagrange multipliers, thus an increasing dimension of the Lagrange multipliers block ($m$) in the augmented system \eqref{eq:aug-alg-sys}. We consider the star-shape domain depicted in Figure \ref{fig::starShape_domain}, which comprises six cylinders with radius $1\ mm$ and length $10\ mm$ connected to a sphere with radius $2\ mm$.
We impose a homogeneous Dirichlet condition at the wall, a homogeneous Neumann condition at $\Gamma^N$, an inflow condition at $\Gamma^{in}$ prescribed with a Lagrange multiplier, and four outflow conditions at $\Gamma^{out}_i$, $i=1,\ldots,4$, prescribed either with Lagrange multipliers or assuming a parabolic Dirichlet profile. For the latter cases, we impose the time-evolving flow rates depicted in Figure \ref{fig::ramp_in-out}. We perform several numerical experiments varying the number of Lagrange multipliers by employing the different approaches for the prescription of the flow rate conditions. We consider a mesh size $h\approx0.2\ mm$, resulting in $\num{141720}$ velocity d.o.f.'s and $\num{47240}$ pressure d.o.f.'s, and a time-step $\Delta t=0.01\ s$. 

\begin{figure}[hbt]
    \centering
    \includegraphics[width=0.45\textwidth]{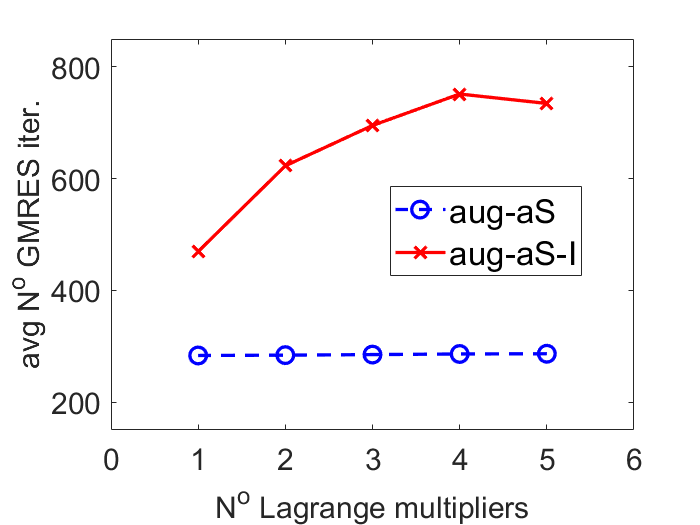}
    \includegraphics[width=0.45\textwidth]{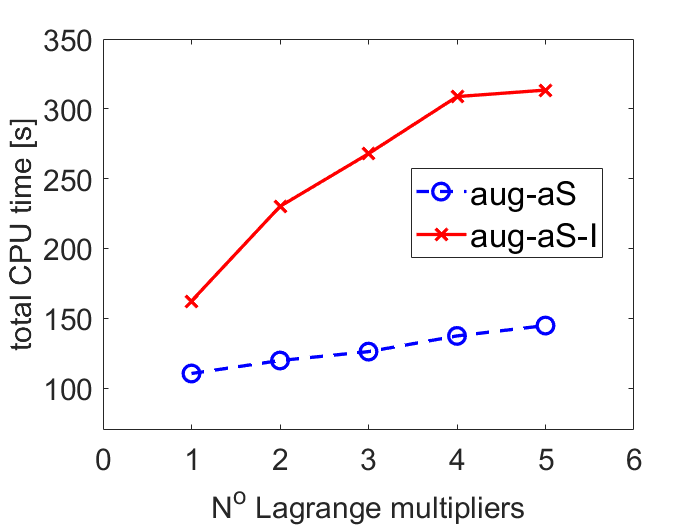}
    \caption{Numerical simulations were run in the star-shape domain (see Figure \ref{fig::starShape_domain}) varying the number of Lagrange multipliers. For both preconditioners aug-aS and aug-aS-I we report as a function of the number of Lagrange multipliers: the average-in-time number of GMRES iterations (left) and the total CPU time (right). Test I.}\label{fig::aSIMPLE_optimality}
\end{figure}

In Figure \ref{fig::aSIMPLE_optimality} we report the average-in-time number of GMRES iterations and the total CPU time as a function of the number of Lagrange multipliers $m$. 
The figure shows that by using the simplified preconditioner aug-aS-I both the analyzed quantities tend to significantly increase with the number of Lagrange multipliers. On the other hand, by using the complete preconditioner aug-aS, we notice that the average number of GMRES iterations does not depend on the size of the Lagrange multipliers block, providing an "optimality" result for the proposed extension. Moreover, the total CPU time slightly increases with the number of Lagrange multipliers, due to the increasing time needed to setup and assemble the linear system. Nonetheless, we can always notice a significant speed-up by using preconditioner aug-aS instead of aug-aS-I.

\subsection{Real-life hemodynamic applications}
\label{subsec:real-life-appl}

We aim to test the numerical performance of preconditioners aug-aS and aug-aS-I in real-life hemodynamic scenarios. In particular, we perform blood-dynamics numerical simulations in patient-specific geometries, representing three different districts of the human body: carotid arteries (Test II), coronary arteries (Test III) and thoracic aorta (Test IV).

We notice that, in accordance to the augmented problem presented in Section \ref{sec:aug-NS-prob}, we always impose a homogeneous Dirichlet condition at the lateral wall (physical boundary) of the computational domains. Moreover, at least at one artificial (not physical) boundary we prescribe a Neumann condition. Here one should prescribe a physiological pressure. However, in the rigid wall context considered in this paper, the pressure is defined up to a constant. For this reason on such boundaries we consider homogeneous Neumann conditions. 

All the numerical simulations were run in parallel on 56 cores of Xeon E5-2640v4@2.4 GHz CPU’s, using the computational resources available at MOX, Dipartimento di Matematica, Politecnico di Milano, with a time-step $\Delta t=0.001\ s$.

\begin{figure}[hbt]
    \centering
    \hspace*{1cm}
    \begin{subfigure}[b]{0.2\textwidth}
        \centering
        \includegraphics[width=\linewidth]{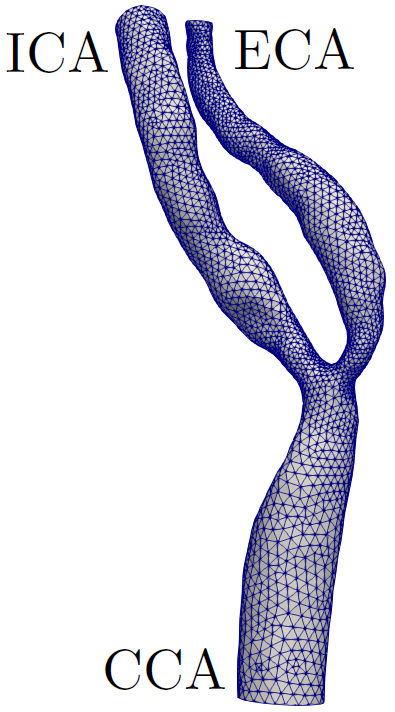}
        \caption{}\label{fig:carotid_domain}
    \end{subfigure}
    \hspace{0.5cm}
    \begin{subfigure}[b]{0.45\textwidth}
        \centering
        \includegraphics[width=\linewidth]{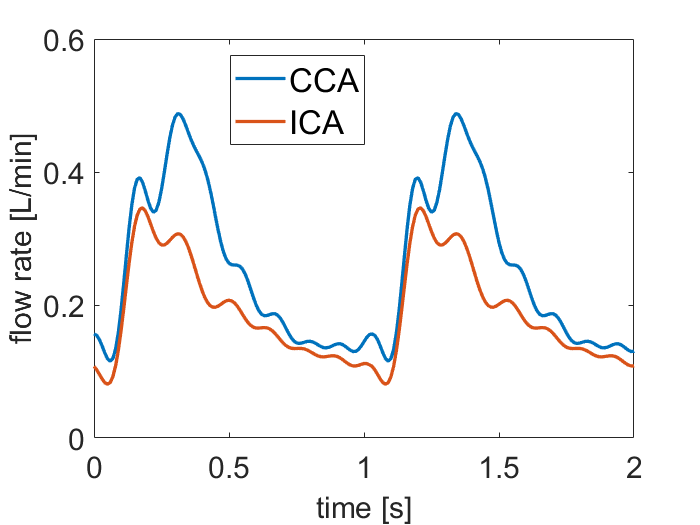}
        \caption{}\label{fig:flow_CCA-ICA}
    \end{subfigure}
    \vskip\baselineskip
    \begin{subfigure}[b]{0.27\textwidth}
        \centering
        \includegraphics[width=\linewidth]{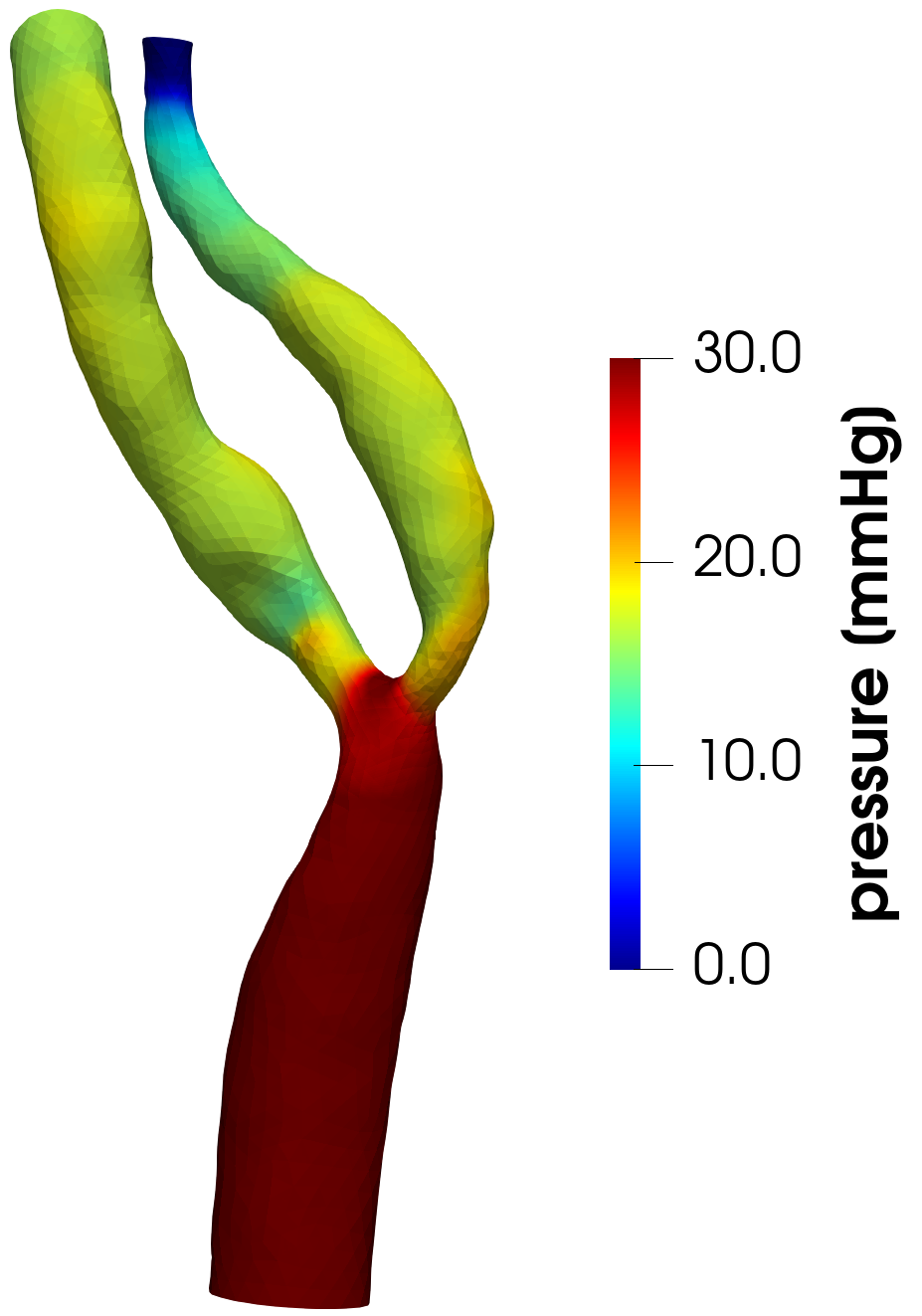}
        \caption{}\label{fig:carotid_pressure}
    \end{subfigure}
    \hspace{1cm}
    \begin{subfigure}[b]{0.3\textwidth}
        \centering
        \includegraphics[width=\linewidth]{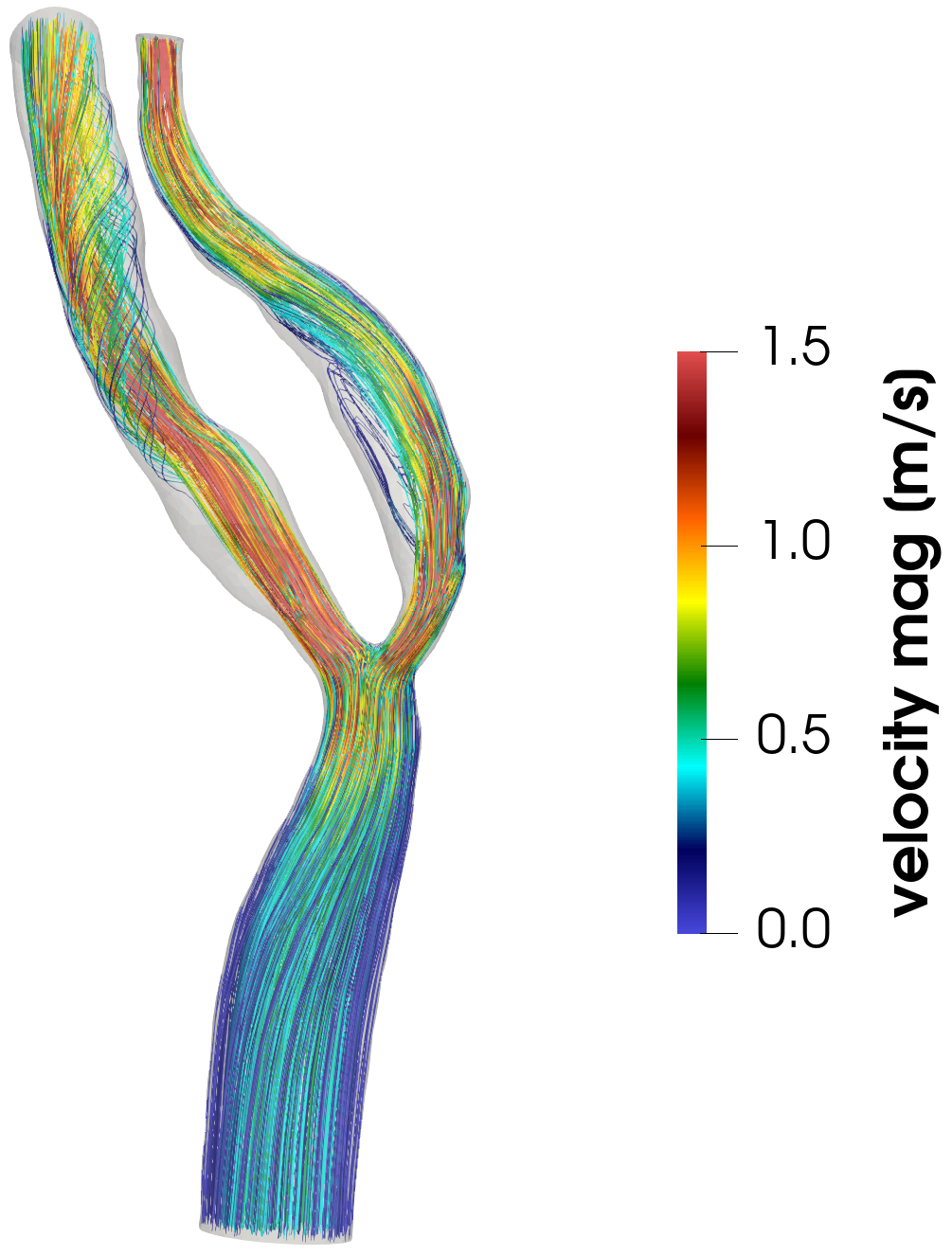}
        \caption{}\label{fig:carotid_velocity}
    \end{subfigure}
    \caption{(a) Carotid arteries computational domain (average mesh size $h\approx0.5\ mm$). (b) Patient-specific flow rate measurements in CCA and ICA, prescribed as inflow at the CCA boundary and as outflow at the ICA boundary. (c)-(d) At the time-instant of peak flow rate in the CCA we report: the pressure distribution (c) and the velocity streamlines, colored by velocity magnitude, (d) obtained in the numerical simulation. Test II.}\label{fig:carotid_panel}
\end{figure}

In the forthcoming three paragraphs we qualitatively describe the numerical results obtained in each real-life scenario, whereas in the last paragraph of the section we discuss the performance of our preconditioners in the three numerical tests together.
\vspace{0.3cm}\\
\noindent\textbf{Test II - Carotid arteries.} We consider a computational domain, taken from \cite{guerciotti2016computational}, representing a patient-specific stenotic carotid artery bifurcation, made of the Common Carotid Artery (CCA), the External Carotid Artery (ECA) and the Internal Carotid Artery (ICA) (see Figure \ref{fig:carotid_domain}).
The average mesh size is $h\approx0.5\ mm$ with a local refinement close to the stenosis, resulting in $\num{45555}$ velocity d.o.f.'s and $\num{15185}$ pressure d.o.f.'s.

We have at disposal patient-specific flow rate measurements in the CCA and the ICA \cite{guerciotti2016computational} (see Figure \ref{fig:flow_CCA-ICA}) and we use them to prescribe an inflow condition at the CCA boundary and an outflow condition at the ICA boundary using two Lagrange multipliers. Moreover, we close the problem by imposing a homogeneous Neumann condition at the ECA boundary.

\begin{figure}[hbt]
    \centering
    \begin{subfigure}[b]{0.4\textwidth}
        \centering
        \includegraphics[width=\linewidth]{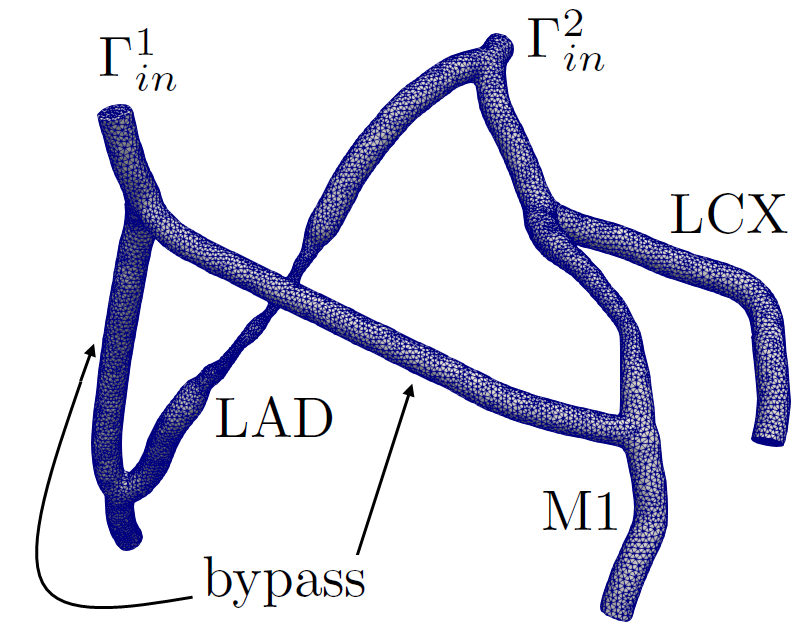}
        \caption{}\label{fig:coronaries_domain}
    \end{subfigure}
    \hspace{0.5cm}
    \begin{subfigure}[b]{0.4\textwidth}
        \centering
        \includegraphics[width=\linewidth]{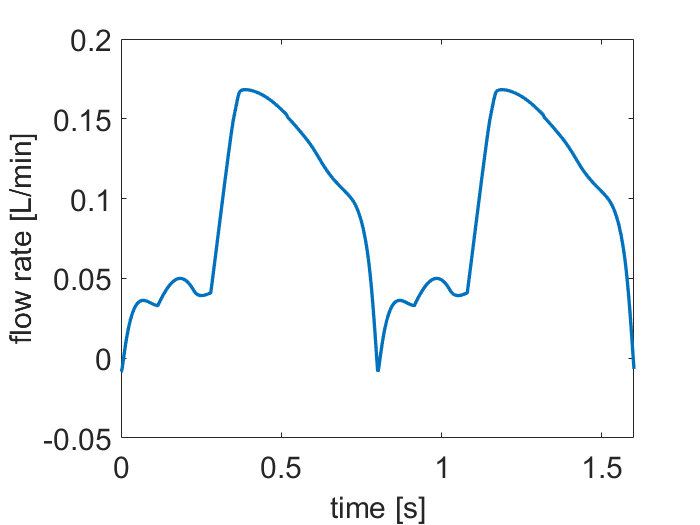}
        \caption{}\label{fig:flow_coronaries}
    \end{subfigure}
    \vskip\baselineskip
    \hspace*{0.2cm}
    \begin{subfigure}[b]{0.4\textwidth}
        \centering
        \includegraphics[width=\linewidth]{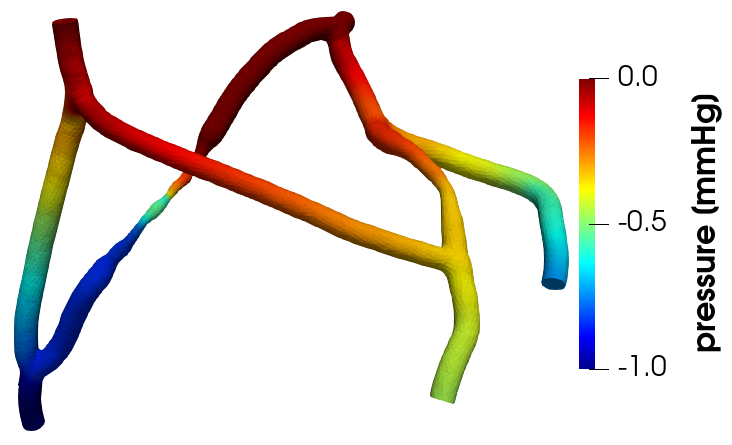}
        \caption{}\label{fig:coronaries_pressure}
    \end{subfigure}
    \hspace{0.5cm}
    \begin{subfigure}[b]{0.4\textwidth}
        \centering
        \includegraphics[width=\linewidth]{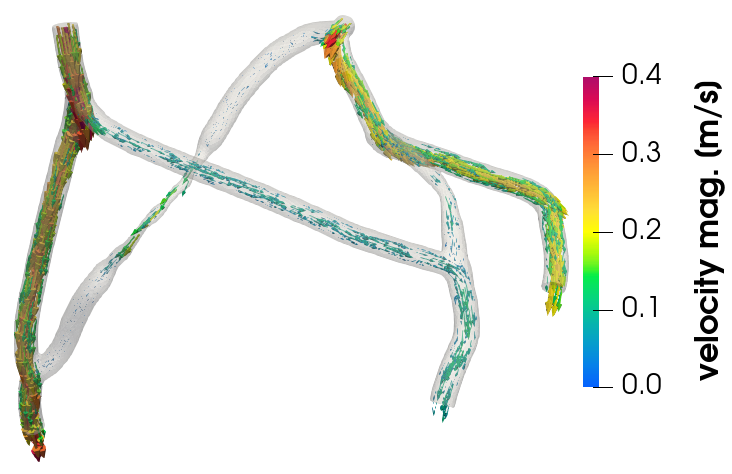}
        \caption{}\label{fig:coronaries_velocity}
    \end{subfigure}
    \caption{(a) Coronary arteries domain comprising three left coronary arteries (LAD, M1 and LCX) and a double bypass originating from the boundary $\Gamma_{in}^1$ ($h\approx0.6\ mm$). (b) Physiological flow rate in left coronary tree \cite{digregorio2022prediction}, partitioned between the three outlet boundaries. (c)-(d) At the time-instant of peak flow rate in the left coronary tree we report: the pressure distribution (c) and the velocity streamlines, colored by velocity magnitude, (d) obtained in the numerical simulation. Test III.}\label{fig:coronaries_panel}
\end{figure}

In Figure \ref{fig:carotid_pressure}-\ref{fig:carotid_velocity} we report the numerical pressure distribution and velocity streamlines at the time-instant of peak flow rate in the CCA. The figure shows a significant pressure drop at the level of the bifurcation, due to a severe narrowing of the cross-sectional area of the domain. Such a narrowing gives rise to high-velocity jets at the bifurcation level and the formation of vortical structures in the bifurcating branches (ICA and ECA) due to another sudden change of the domain's morphology.
\vspace{0.3cm}\\
\noindent\textbf{Test III - Coronary arteries.} We consider a patient-specific post-surgical geometry, taken from \cite{guerciotti2017computational}, representing a district of the left coronary tree comprising three coronary arteries (LAD, M1 and LCX) and a double bypass. A bypass creates a new path for blood to flow around a partially blocked (stenotic) artery restoring a normal blood flow downstream of the blockage (stenosis). In our case, a double bypass, originating from the boundary $\Gamma_{in}^1$, was needed to treat the patient since both the LAD and M1 arteries present stenoses (see Figure \ref{fig:coronaries_domain}).
The average mesh size is $h\approx0.6\ mm$ with a local refinement close to the stenoses, resulting in $\num{90015}$ velocity d.o.f.'s and $\num{30005}$ pressure d.o.f.'s.

The domain presents two inlet sections, the bypass inlet $\Gamma^1_{in}$ and the left coronary artery inlet $\Gamma^2_{in}$, where we impose homogeneous Neumann conditions, thus assuming to have the same pressure upstream the stenoses. Moreover, we guarantee the physiological flow rate taken from \cite{digregorio2022prediction} (see Figure \ref{fig:flow_coronaries}) inside the left coronary tree by imposing three outflow conditions, with three Lagrange multipliers, which partition the total flow in such a way that it goes: $\frac{1}{6}$ through the M1 outlet, $\frac{1}{3}$ through the LCX outlet, and $\frac{1}{2}$ through the remaining outlet. By doing so we are assuming that the double bypass restores a physiological flow downstream the stenoses.

\begin{figure}[hbt]
    \centering
    \hspace*{0.6cm}
    \begin{subfigure}[b]{0.3\textwidth}
        \centering
        \includegraphics[width=\linewidth]{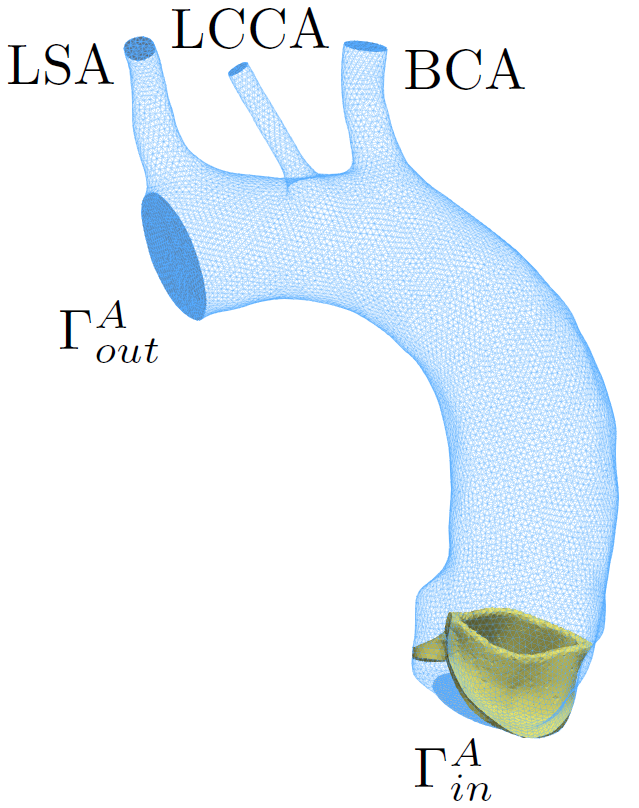}
        \caption{}\label{fig:aorta_domain}
    \end{subfigure}
    \hspace{1cm}
    \begin{subfigure}[b]{0.45\textwidth}
        \centering
        \includegraphics[width=\linewidth]{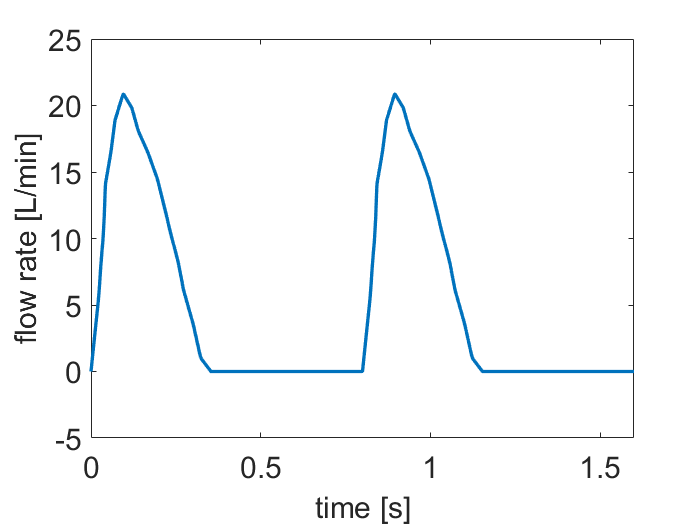}
        \caption{}\label{fig:flow_aorta}
    \end{subfigure}
    \vskip\baselineskip
    \begin{subfigure}[b]{0.4\textwidth}
        \centering
        \includegraphics[width=\linewidth]{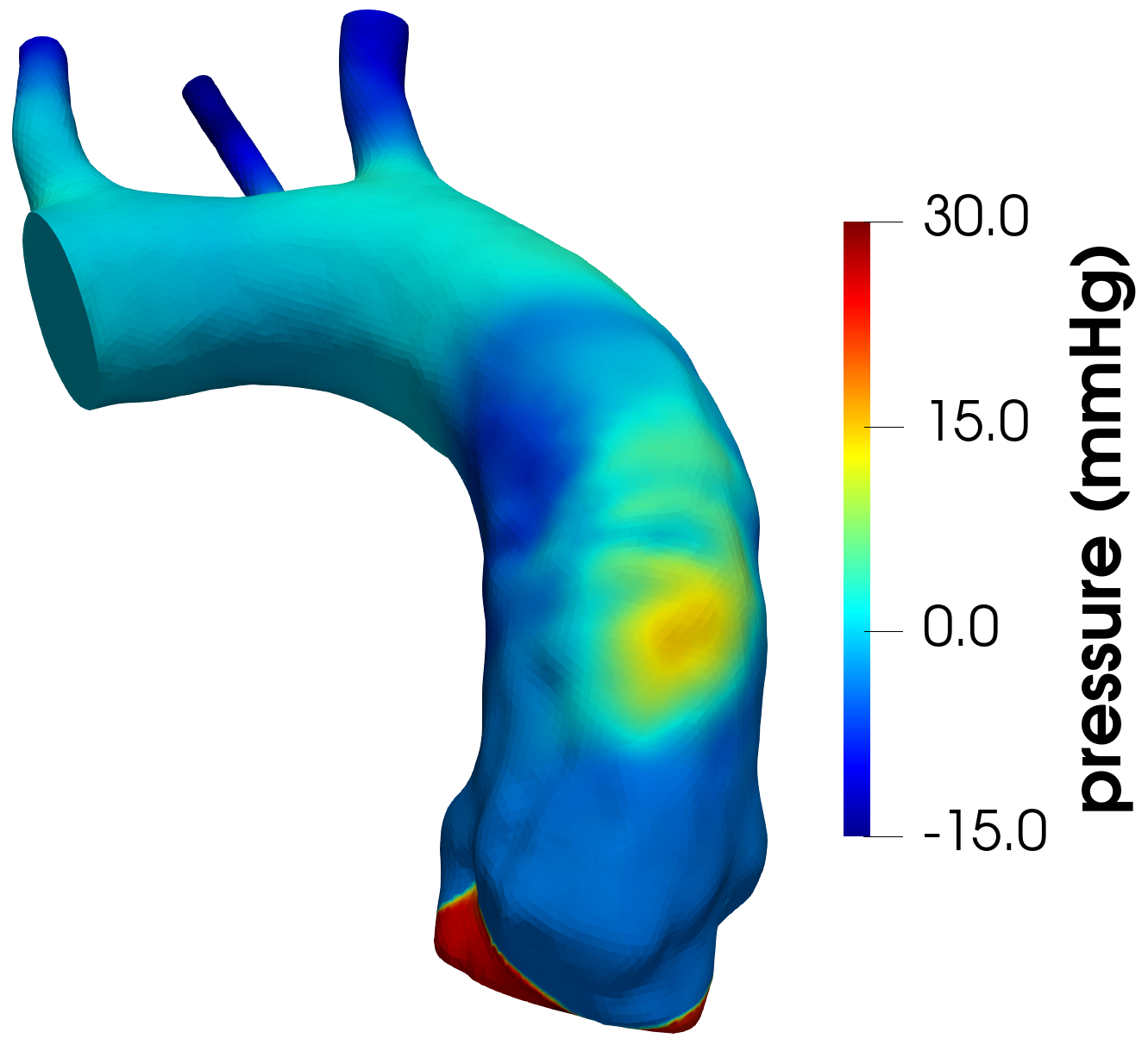}
        \caption{}\label{fig:aorta_pressure}
    \end{subfigure}
    \hspace{0.5cm}
    \begin{subfigure}[b]{0.4\textwidth}
        \centering
        \includegraphics[width=\linewidth]{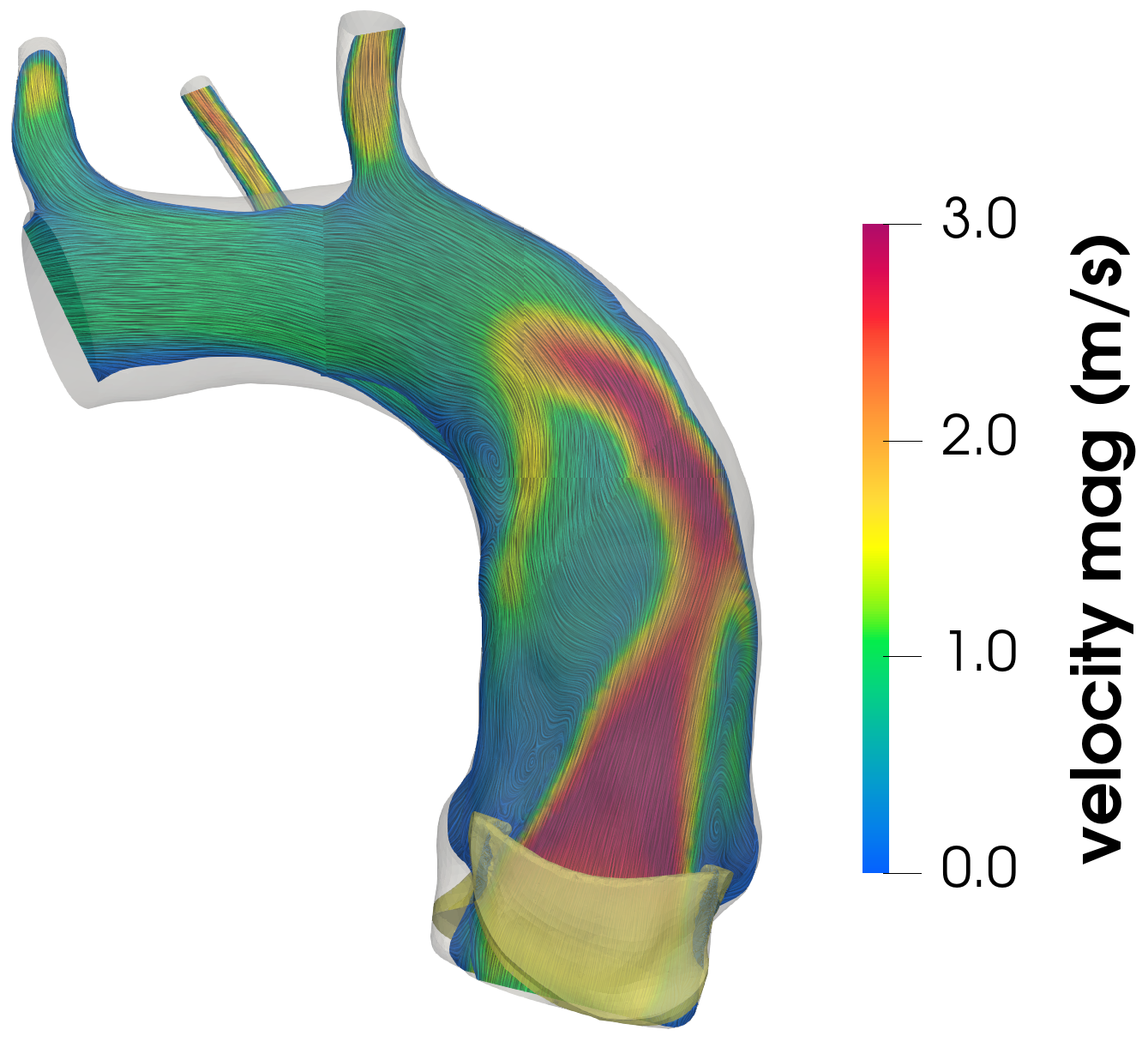}
        \caption{}\label{fig:aorta_velocity}
    \end{subfigure}
    \caption{(a) Thoracic aorta domain with the bicuspid aortic valve's leaflets in yellow ($h\approx1.3\ mm$). (b) Physiological flow rate \cite{capellini2018computational} prescribed at $\Gamma_{in}^A$ and partitioned between the BCA, LCCA and LSA outlet boundaries. (c)-(d) At the time-instant of peak flow rate in the thoracic aorta we report: the pressure distribution (c) and the velocity streamlines, colored by velocity magnitude, on a longitudinal section (d) obtained in the numerical simulation. Test IV.}\label{fig:aorta_panel}
\end{figure}

In Figure \ref{fig:coronaries_pressure}-\ref{fig:coronaries_velocity} we report the numerical pressure distribution and velocity streamlines at the time-instant of peak flow rate in the left coronary tree.
From the velocity visualization we see that the majority of the flow is passing through the bypasses and not through the stenotic tracts of the LAD and M1 arteries. This results in a moderate pressure gradient, around $1\ mmHg$, between inlet and outlet sections, thus avoiding significant pressure losses across the stenoses which would be present without the bypasses.
\vspace{0.3cm}\\
\noindent\textbf{Test IV - Thoracic aorta.} We consider a patient-specific aortic domain, taken from \cite{bonomi2015influence}, going from the aortic root to the aortic arch, including the three sovra-aortic branches. In particular, the patient under consideration has a native bicuspid aortic valve, which differs from a normal aortic valve for the presence of two valve's leaflets instead of three (see Figure \ref{fig:aorta_domain}).
The aortic valve's leaflets in their open configuration are implicitly represented in the numerical simulation as an obstruction to the flow using the Resistive Implicit Immersed Surface (RIIS) method \cite{fernandez2008numerical,astorino2012robust,fedele2017patient,fumagalli2020image}. This consists in adding a local penalization term inside the momentum equation of the Navier-Stokes system to model the adherence of blood to the leaflets. We consider also the Large Eddy Simulation (LES) $\sigma$-model \cite{nicoud2011using,lancellotti2017large} to account for the presence of turbulence inside the aorta \cite{stein1976turbulent}. Moreover, in order to prevent numerical instabilities, we add a backflow stabilization, as proposed in \cite{esmaily2011comparison}, at both the aortic inlet $\Gamma_{in}^A$ and the aortic outlet $\Gamma_{out}^A$. RIIS, LES and backflow stabilization introduce specific changes in the first block row of system \eqref{eq:aug-alg-sys} \textcolor{black}{(see Appendix \ref{app:TestIV-algebraic}, where the value of specific parameters are also reported)}. In this case, preconditioners \eqref{eq:aug-aSIMPLE-inexact} and \eqref{eq:aug-aSIMPLE-identity} are simply adapted by considering such changes without any approximations. The average mesh size is $h\approx1.3\ mm$ with a local refinement of $0.4\ mm$ in the leaflets area, resulting in $\num{377673}$ velocity d.o.f.'s and $\num{125891}$ pressure d.o.f.'s.

We impose an inflow condition at $\Gamma_{in}^A$, prescribing the physiological flow rate taken from \cite{capellini2018computational} (see Figure \ref{fig:flow_aorta}) entering the aorta, and three outflow conditions at the BCA, LCCA and LSA boundaries, guaranteeing that $30\%$ of the total flow goes to these sovra-aortic arteries. Specifically, we split the flow between each sovra-aortic branch according to the area of its outlet boundary. As a result, the inlet flow is partitioned in such a way that it goes: $17.0\%$ through the BCA outlet, $4.8\%$ through the LCCA outlet, and $8.2\%$ through the LSA outlet. All these flow rate conditions are prescribed using four Lagrange multipliers. Moreover, we close the problem by imposing a homogeneous Neumann condition at $\Gamma_{out}^A$.

In Figure \ref{fig:aorta_pressure}-\ref{fig:aorta_velocity} we report the numerical pressure distribution in the domain and velocity streamlines on a longitudinal section at the time-instant of peak flow rate in the aorta. Bicuspid aortic valves are characterized by a smaller orifice area with respect to the normal tricuspid valves. In our numerical results this translates in a significant pressure drop across the valve and in the formation of a high-velocity jet inside the first tract of the aorta. This jet collides against the aortic wall giving rise to high pressure areas on the wall. Moreover, we observe the formation of vortical structures along side the jet, giving rise to complex hemodynamic patterns in the aorta.
\vspace{0.3cm}\\
\noindent \textbf{Performance of the augmented aSIMPLE preconditioner.} Let us first stress that the solutions reported in Figure \ref{fig:carotid_panel}-\ref{fig:coronaries_panel}-\ref{fig:aorta_panel} are the same (up to the GMRES tolerance) for the two preconditioners aug-aS and aug-aS-I, since we are in both cases at convergence. In the following we discuss the performance in terms of efficiency of the two preconditioners.

\begin{figure}[htb]
    \centering
    \begin{subfigure}[b]{0.32\textwidth}
        \centering
        \includegraphics[width=\linewidth]{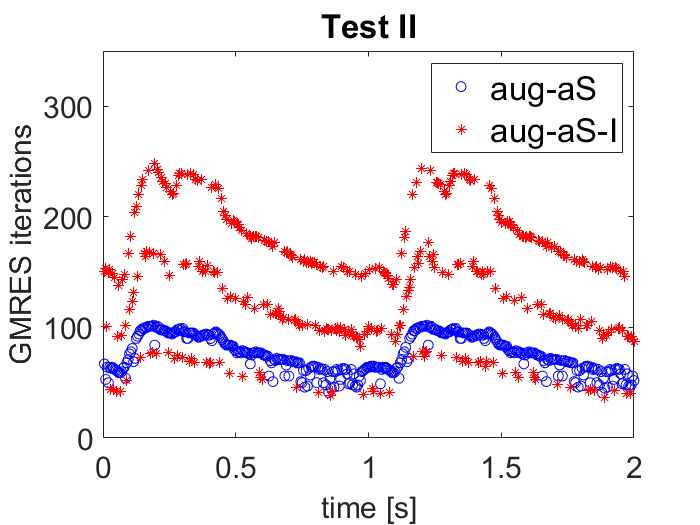}
        \caption{}\label{fig:carotid_GMRESiter}
    \end{subfigure}
    \begin{subfigure}[b]{0.32\textwidth}
        \centering
        \includegraphics[width=\linewidth]{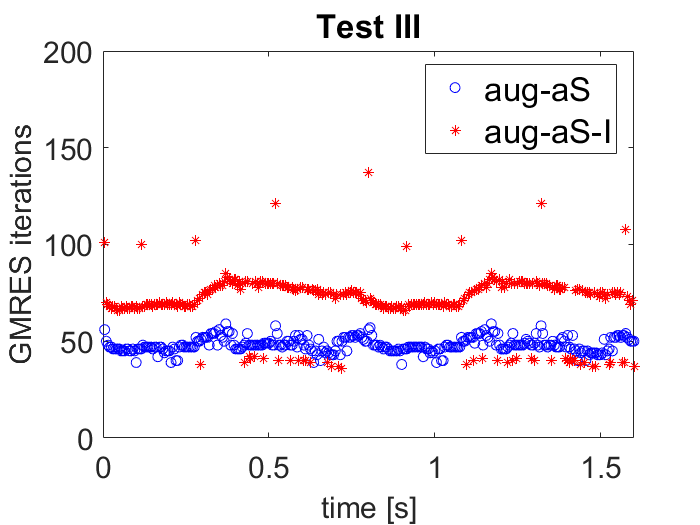}
        \caption{}\label{fig:coronaries_GMRESiter}
    \end{subfigure}
    \begin{subfigure}[b]{0.32\textwidth}
        \centering
        \includegraphics[width=\linewidth]{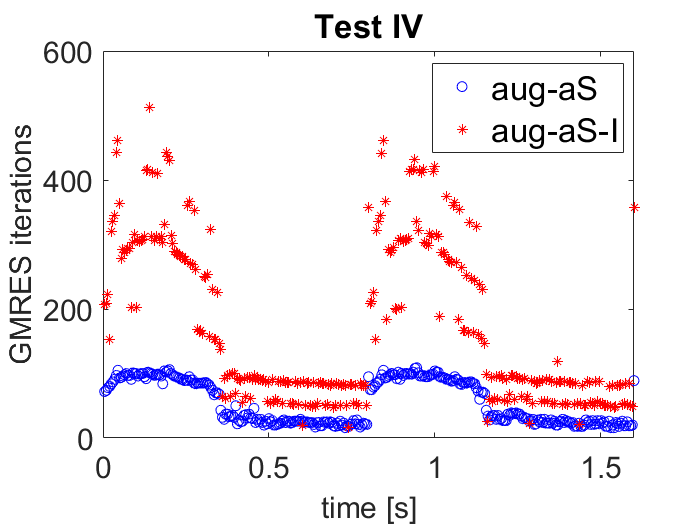}
        \caption{}\label{fig:aorta_GMRESiter}
    \end{subfigure}
    \caption{Number of GMRES iterations as a function of time needed to monolithically solve each real-life hemodynamic problem using preconditioner aug-aS (blue) and aug-aS-I (red). Notice that, in each sub-plot, the red stars are all related to one numerical simulation.}\label{fig:GMRESiter_compare}
\end{figure}

In Table \ref{tab:avgGMRESiter} we report the average-in-time number of GMRES iterations needed to monolithically solve each real-life hemodynamic problem presented above using both the complete and simplified preconditioners. 
\begin{table}[hbt]
    \caption{Average-in-time number of GMRES iterations in each real-life hemodynamic application using preconditioners aug-aS and aug-aS-I.}
    \centering
    \renewcommand{\arraystretch}{1.5}
    \begin{tabular}{|c|c|c|}
        \hline
        & aug-aS & aug-aS-I \\
        \hline
        Test II - Carotid & $75.8$ & $150.6$ \\
        Test III - Coronary & $48.0$ & $70.8$ \\
        Test IV - Aorta & $55.3$ & $175.7$ \\
        \hline
    \end{tabular}
    \label{tab:avgGMRESiter}
\end{table}
The table shows that in all the applications the average number of GMRES iterations for the aug-aS case is much smaller than for the aug-aS-I case. This is also confirmed by the results reported in Figure \ref{fig:GMRESiter_compare}, where the number of GMRES iterations as a function of time is shown.
Indeed, we can see that at almost each time-step the number of iterations associated to aug-aS is smaller than the one associated to aug-aS-I.

We can also observe that in all the numerical experiments considering preconditioner aug-aS-I\footnote{Notice that, in each sub-plot, the red stars are all related to one numerical simulation} the number of GMRES iterations is highly variable and presents severe oscillations during the time evolution; such behavior is not present for preconditioner aug-aS. This highlights the better stability of aug-aS due to the proper approximation of the blocks related to the Lagrange multipliers (third row in L and third column in U in the block LU factorization \eqref{eq:aug-exact-LU}).

Moreover, from Figure \ref{fig:GMRESiter_compare} it emerges that for both preconditioners the number of iterations is higher in the time-instants characterized by a high flow rate. This can be explained by the fact that the diagonal approximation does not capture enough information about the convection operator \cite{elman2008taxonomy}, thus poor performance of the SIMPLE preconditioner, which is based on such approximation, are to be expected in a convection-dominated regime. In particular, this behavior is mitigated for preconditioner aug-aS. 

\begin{table}[hbt]
    \caption{Application: real-life hemodynamic application; $Re^{max}$: maximum Reynolds number in the simulation; $N_{points}^{max}$: maximum number of mesh points among the boundaries where flow rate conditions are prescribed using Lagrange multipliers; $N_{LagMult}$: number of Lagrange multipliers considered in the numerical simulation; $d_{\text{GMRES}}^{\%}$: relative difference between the average number of GMRES iterations obtained using preconditioner aug-aS with respect to preconditioner aug-aS-I; $d_{\text{CPU}}^{\%}$: relative difference between the total CPU time using preconditioner aug-aS with respect to preconditioner aug-aS-I.}
    \centering
    \renewcommand{\arraystretch}{2}
    \begin{tabular}{|c|c|c|c|c|c|}
        \hline
        Application & $Re^{max}$ & $N_{points}^{max}$ & $N_{LagMult}$ & $d_{\text{GMRES}}^{\%}$ & $d_{\text{CPU}}^{\%}$  \\
        \hline
        Carotid & $554$ & $70$ & $2$ & $-49.67\%$ & $-33.11\%$ \\
        Coronary & $184$ & $36$ & $3$ & $-32.16\%$ & $-4.03\%$ \\
        Aorta & $4730$ & $408$ & $4$ & $-68.54\%$ & $-28.16\%$ \\
        \hline
    \end{tabular}
    \label{tab:recap}
\end{table}

The results reported in Table \ref{tab:recap} suggest that it is particularly significant to employ preconditioner aug-aS when the boundaries at which flow rate conditions are prescribed using Lagrange multipliers contain a great number of mesh points (represented in the table by the maximum number of mesh points $N_{points}^{max}$ among these boundaries). Indeed, the percentage of saving in the number of GMRES iterations $d_{\text{GMRES}}^{\%}$ increases with $N_{points}^{max}$. This is due to matrix $\Phi$ in \eqref{eq:aug-alg-matrix} which has several non-zero entries when $N_{points}^{max}$ is large and thus aug-aS-I preconditioner introduces in this case a too rough approximation. 

From the same table we also observe that the dependence on the number of Lagrange multipliers $N_{LagMult}$ of the speed-up in terms of GMRES iterations and CPU time allowed by the use of the complete preconditioner aug-aS seems to not have a clear trend.

To conclude, the proposed trivial extension of the standard aSIMPLE preconditioner aug-aS-I can be suited in applications where $Re^{max}$ and $N_{points}^{max}$ are small, e.g. Test III which shows a speed-up in terms of CPU time of just $4.03\%$ (against much larger values in the other cases, see last column of Table \ref{tab:recap}
) by using preconditioner aug-aS instead of aug-aS-I. In all other applications the complete extension of the aSIMPLE preconditioner aug-aS should be preferred.

\subsection{On the Dirichlet approach for flow rate conditions}
\label{subsec:Def-vs-Dir}

\begin{figure}[hbt]
    \centering
    \begin{subfigure}[b]{0.4\textwidth}
        \centering
        \includegraphics[width=\linewidth]{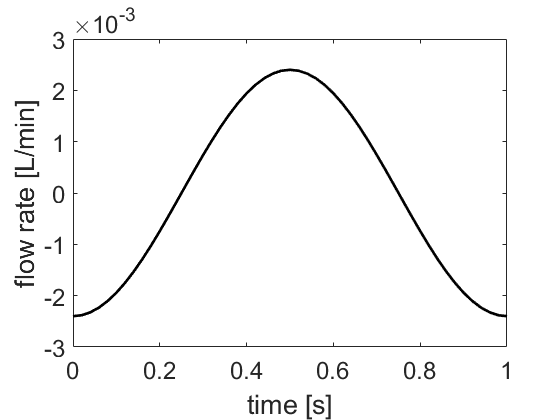}
        \caption{}\label{fig::pulsatileFlowrate}
    \end{subfigure}
    \hspace{0.2cm}
    \begin{subfigure}[b]{0.55\textwidth}
        \centering
        \includegraphics[width=\linewidth]{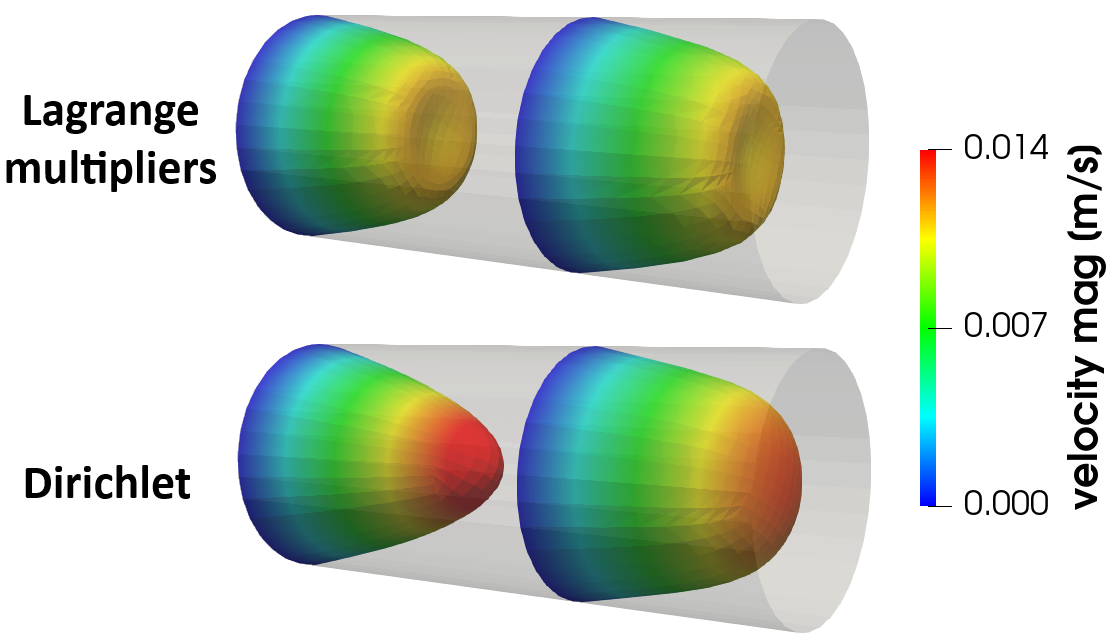}
        \caption{}\label{fig::LagMult_vs_Dir_cyl}
    \end{subfigure}
    \caption{(a) Temporal evolution of the sinusoidal pulsatile flow rate imposed at the inlet boundary of the cylinder. (b) Numerical velocity profile, colored by velocity magnitude, in the cylindrical domain at $t=0.5\ s$ (peak flow rate time-instant) for both the Lagrange multipliers and Dirichlet treatment of the inlet flow rate condition. Notice that the cylinder has been stretched along its longitudinal direction for visualization purposes.}\label{fig::inlet_Womer_profiles}
\end{figure}

In this subsection we highlight the main advantages of treating defective flow rate conditions following the Lagrange multipliers approach instead of prescribing a selected Dirichlet condition. In particular, we consider a toy problem and a real-life application.
\vspace{0.3cm}\\
\noindent\textbf{Toy problem:} Let us consider a cylindrical domain (radius $4\ mm$) and impose a homogeneous Dirichlet condition at the lateral wall (no-slip), a homogeneous Neumann condition at the outlet section and a flow rate condition at the inlet section, prescribing the pulsatile flow rate in Figure \ref{fig::pulsatileFlowrate}. We are interested in studying the velocity spatial profile obtained by employing two different approaches for the prescription of the inlet condition: using a Lagrange multiplier or assigning a Dirichlet condition, assuming a parabolic profile. Being an axi-symmetric cylindrical domain subject to a sinusoidal periodic flow rate, there is an analytical solution given by the Womersley solution \cite{womersley1955method,berselli2013exact}. In particular, it was already proven in \cite{veneziani2005flow} that the numerical solution obtained with the augmented formulation \eqref{eq:aug-var-form} matches the analytical one. We consider an average mesh size $h\approx0.7\ mm$, resulting in $\num{41451}$ velocity d.o.f.'s and $\num{13817}$ pressure d.o.f.'s, and a time-step $\Delta t=0.001\ s$

In Figure \ref{fig::LagMult_vs_Dir_cyl} we show, for both the Lagrange multipliers and the Dirichlet approach, the numerical velocity spatial profiles on two different cross-sections (one at the inlet and one inside the domain) at the peak flow rate time-instant. Since in \cite{veneziani2005flow} has been shown that the augmented formulation is able to perfectly match the Womersley analytical solution, the plot at the top of Figure \ref{fig::LagMult_vs_Dir_cyl} displays the exact/reference solution. By observing the Dirichlet solution reported at the bottom we notice that the velocity profile differs from the reference solution also inside the domain, due to the influence of the parabolic profile prescribed at the inlet. This shows the need to use Lagrange multipliers to achieve accurate numerical solutions also in a regular geometry.
\vspace{0.3cm}\\
\noindent\textbf{Real-life application:} We consider the thoracic aorta problem introduced in Section \ref{subsec:real-life-appl}, neglecting the presence of the aortic valve's leaflets. In particular, we compare the numerical solutions obtained by imposing the inlet flow rate reported in Figure \ref{fig:flow_aorta} using a Lagrange multiplier and using a Dirichlet condition, assuming a flat velocity spatial profile (as commonly done in the aortic root \cite{rossvoll1991velocity}). 

\begin{figure}[hbt]
    \centering
    \begin{subfigure}[b]{0.42\textwidth}
        \centering
        \includegraphics[width=\linewidth]{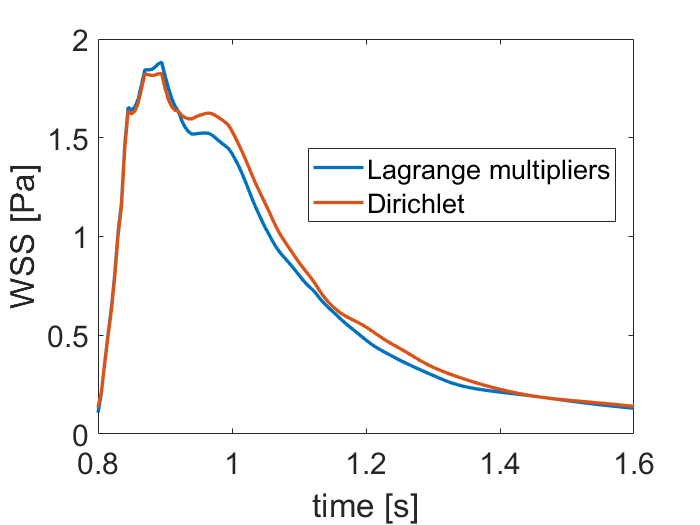}
        \caption{}\label{fig::avgWSS_LagMult-Dir}
    \end{subfigure}
    \begin{subfigure}[b]{0.53\textwidth}
        \centering
        \includegraphics[width=\linewidth]{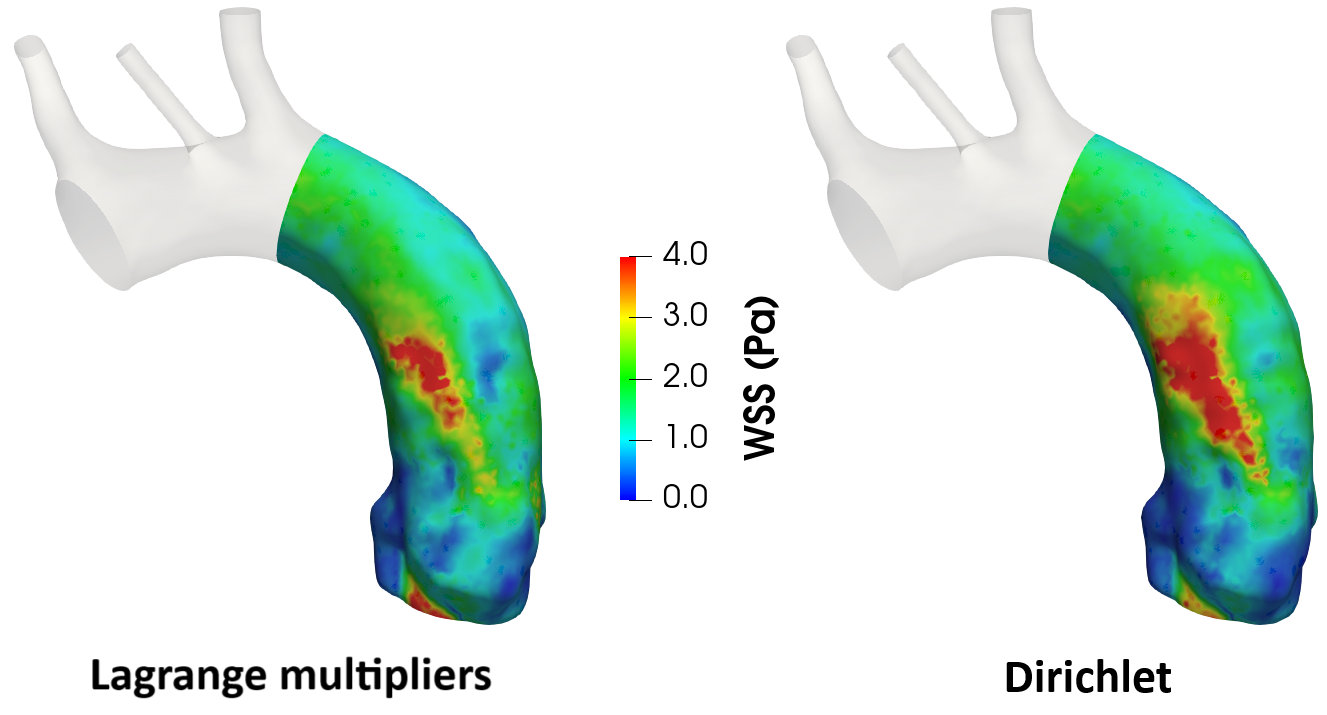}
        \caption{}\label{fig::pointWSS_LagMult-Dir}
    \end{subfigure}
    \caption{For both the Lagrange multipliers and Dirichlet treatment of the inlet flow rate condition we report: the temporal evolution of the average WSS on the aortic wall (a) and the WSS pattern on the aortic wall at $t=1\ s$ (b). }\label{fig::aorta_WSS}
\end{figure}

In Figure \ref{fig::avgWSS_LagMult-Dir} we report, only for the second cycle, the temporal evolution of the average Wall Shear Stress (WSS) on the aortic wall for both inlet condition treatments.  
During the first part of the simulation, when the inlet flow rate is increasing (see Figure \ref{fig:flow_aorta}), the two numerical solutions coincide, suggesting that the flat profile assumption is accurate in this period of the cycle. On the other, after this acceleration phase, the Dirichlet solution starts showing non-negligible differences with respect to the Lagrange multipliers one. This can be explained by a change of the velocity spatial profile in the aortic root, which we are able to capture only following the Lagrange multipliers approach. 

This is confirmed by the results reported in Figure \ref{fig::pointWSS_LagMult-Dir}, where we depict the WSS pattern on the aortic wall at a time-instant after the acceleration phase ($t=1\ s$), which highlights significance difference among the two solutions.

We conclude that imposing a flow rate by prescribing a Dirichlet boundary condition can influence the accuracy of the numerical solution even far away from the boundary.

\appendix
\section{Algebraic system for Test IV - Thoracic aorta}\label{app:TestIV-algebraic}
In this appendix we provide additional details on the RIIS, LES $\sigma$-model and backflow stabilization methods considered in "Test IV - Thoracic aorta" (see Section \ref{subsec:real-life-appl}). Specifically, we will focus on how these methods change the augmented variational formulation \eqref{eq:aug-var-form}, the corresponding augmented monolithic system \eqref{eq:aug-alg-sys}, preconditioner aug-aS \eqref{eq:aug-aSIMPLE-inexact} and preconditioner aug-aS-I \eqref{eq:aug-aSIMPLE-identity}. 

Each of these methods introduces a term in the left hand side of the first equation of the augmented variational formulation \eqref{eq:aug-var-form}. In particular, we follow the same notation introduced in Section \ref{sec:aug-NS-prob} and Section \ref{subsec:real-life-appl}.
\begin{itemize}
    \item[-] RIIS:
        \begin{equation*}
        \displaystyle\left(\delta_{\Sigma,\epsilon}\frac{R}{\epsilon}(\textbf{u}-\textbf{u}_{\Sigma}),\textbf{v}\right),
        \end{equation*}
        where $\Sigma$ is the leaflets surface, $R=10^5\ Kg/m\cdot s$ represents a resistance coefficient, $\epsilon=7\times10^{-4}\ m$ is a numerical parameter representing the half-thickness of a smoothed Dirac delta function $\delta_{\Sigma,\epsilon}$ centered in $\Sigma$ and $\textbf{u}_{\Sigma}=\textbf{0}\ m/s$ is the velocity of the bioprosthetic leaflets that we set to zero since we do not consider leaflets' dynamics \cite{fedele2017patient,bennati2023turbulent}. In Test IV this RIIS term is active only during the ejection phase (positive values of the inlet flow rate in Figure \ref{fig:flow_aorta});
    \item[-] LES $\sigma$-model:
    \begin{equation*}
        \displaystyle\left(\nu_{sgs}(\textbf{u})\nabla\textbf{u},\nabla\textbf{v}\right),
        \end{equation*}
        where $\nu_{sgs}=C\,h^2\,\sigma_3(\sigma_1-\sigma_2)(\sigma_2-\sigma_3)/\sigma_1^2$ (with $\sigma_j$ the singular values of 
        $\nabla\textbf{u}$, $C=1.5$) is a sub-grid scale viscosity which accounts for the non-resolved scales \cite{nicoud2011using};
    \item[-] Backflow stabilization: 
    \begin{equation*}
        \displaystyle-\frac{1}{2}\int_{\Gamma_{in}^A\cup\Gamma_{out}^A}{|\textbf{u}\cdot\textbf{n}|_{-}\textbf{u}\cdot\textbf{v}\ d\gamma},
    \end{equation*}
    where $|\textbf{u}\cdot\textbf{n}|_{-}=\frac{\textbf{u}\cdot\textbf{n}-|\textbf{u}\cdot\textbf{n}|}{2}$, so that this term is different from zero only in the points where $\textbf{u}\cdot\textbf{n}<0$.
\end{itemize}

Considering the same spatial and temporal discretizations presented in Section \ref{sec:aug-NS-prob} with a semi-implicit treatment of the LES $\sigma$-model and backflow stabilization terms, the resulting monolithic system reads:
\begin{equation}\label{eq:alg-aug-sys-test4}
	\begin{bmatrix}
	{K}+G\left(\mathbf{U}^n\right) & B^T & \Phi^T \\
	-B & S & 0 \\
	\Phi & 0 & 0
	\end{bmatrix}
	\begin{bmatrix}
	\bm{U} \\
	\bm{P} \\
	\bm{\Lambda}
	\end{bmatrix}
	=
	\begin{bmatrix}
	{\bm{F}} \\
	\bm{0} \\
	\bm{Q}
	\end{bmatrix},
\end{equation}
showing the same block structure as in \eqref{eq:aug-alg-sys}. The only difference is given by the addition of matrix G in the momentum equation accounting for RIIS, LES, backflow stabilization:
$G(\bm{U}^n) = M_{RIIS}+A_{LES}(\bm{U}^n)+M_{bf}(\bm{U}^n)$, where $[M_{RIIS}]_{kj}=\left(\delta_{\Sigma,\epsilon}\frac{R}{\epsilon}\bm{\psi}_j,\bm{\psi}_k\right)$, $[A_{LES}]_{kj}=\left(\nu_{sgs}(\bm{U}^n)\nabla\bm{\psi}_j,\nabla\bm{\psi}_k\right)$, and \\$[M_{bf}(\bm{U}^n)]_{kj}=-\frac{1}{2}\int_{\Gamma_{in}^A\cup\Gamma_{out}^A}{|\bm{U}^n\cdot\bm{n}|_{-}\bm{\psi}_j\cdot\bm{\psi}_k}\ d\gamma$, $j,k=1,\ldots,N_U$. Preconditioners aug-aS and aug-aS-I are easily obtained from the system matrix in \eqref{eq:alg-aug-sys-test4} by following the same arguments of Sections \ref{subsec:aug-SIMPLE}-\ref{subsec:aug-aSIMPLE}-\ref{subsec:num-exp-set}.

\section*{Acknowledgments}
Luca Crugnola and Christian Vergara are members of the INdAM group GNCS “Gruppo Nazionale per il Calcolo Scientifico” (National Group for Scientific Computing). CV has been partially supported by the Italian Ministry of University and Research (MIUR) within the PRIN (Research projects of relevant national interest) MIUR PRIN22-PNRR n. P20223KSS2 "Machine learning for fluid-structure interaction in cardiovascular problems: efficient solutions, model reduction, inverse problems,and by  the Italian Ministry of Health within the PNC PROGETTO HUB LIFE SCIENCE - DIAGNOSTICA AVANZATA (HLS-DA) "INNOVA", PNC-E3-2022-23683266–CUP: D43C22004930001, within the "Piano Nazionale Complementare Ecosistema Innovativo della Salute” - Codice univoco investimento: PNC-E3-2022-23683266.

\clearpage

\bibliographystyle{siamplain}
%\bibliography{Crugnola-Vergara-aRxiv}

\end{document}